\documentclass[aps,superscriptaddress,twocolumn,driverfallback=dvips]{revtex4}
\usepackage{feynmp}
\usepackage{amsmath}
\usepackage{amssymb}
\usepackage{epsfig}
\usepackage{graphicx}
\usepackage{subfigure}
\usepackage{hyperref}
\usepackage{siunitx}
\usepackage{bbm}
\usepackage{epsfig}
\usepackage{color}
\usepackage{float}
\usepackage{mhchem}
\usepackage{epstopdf}

\newcommand{\Rmnum}[1]{\expandafter\@slowromancap\romannumeral #1@}
\allowdisplaybreaks[3]

\begin{document}

\title{Theoretical study of phonon-mediated superconductivity beyond
Migdal-Eliashberg approximation and Coulomb pseudopotential}

\author{Jie Huang}
\affiliation{Department of Modern Physics, University of Science and
Technology of China, Hefei, Anhui 230026, China}
\author{Zhao-Kun Yang}
\affiliation{Department of Modern Physics, University of Science and
Technology of China, Hefei, Anhui 230026, China}
\author{Xiao-Yin Pan}
\altaffiliation{Corresponding author: panxiaoyin@nbu.edu.cn}
\affiliation{Department of Physics, Ningbo University, Ningbo,
Zhejiang 315211, China}
\author{Guo-Zhu Liu}
\altaffiliation{Corresponding author: gzliu@ustc.edu.cn}
\affiliation{Department of Modern Physics, University of Science and
Technology of China, Hefei, Anhui 230026, China}

\begin{abstract}
In previous theoretical studies of phonon-mediated superconductors,
the electron-phonon coupling is treated by solving the
Migdal-Eliashberg equations under the bare vertex approximation,
whereas the effect of Coulomb repulsion is incorporated by
introducing one single pseudopotential parameter. These two
approximations become unreliable in low carrier-density
superconductors in which the vertex corrections are not small and
the Coulomb interaction is poorly screened. Here, we shall go beyond
these two approximations and employ the Dyson-Schwinger equation
approach to handle the interplay of electron-phonon interaction and
Coulomb interaction in a self-consistent way. We first derive the
exact Dyson-Schwinger integral equation of the full electron
propagator. Such an equation contains several unknown
single-particle propagators and fermion-boson vertex functions, and
thus seems to be intractable. To solve this difficulty, we further
derive a number of identities satisfied by all the relevant
propagators and vertex functions and then use these identities to
show that the exact Dyson-Schwinger equation of electron propagator
is actually self-closed. This self-closed equation takes into
account not only all the vertex corrections, but also the mutual
influence between electron-phonon interaction and Coulomb
interaction. Solving it by using proper numerical methods leads to
the superconducting temperature $T_{c}$ and other quantities. As an
application of the approach, we compute the $T_{c}$ of the
interfacial superconductivity realized in the one-unit-cell
FeSe/SrTiO$_{3}$ system. We find that $T_{c}$ can be strongly
influenced by the vertex corrections and the competition between
phonon-mediated attraction and Coulomb repulsion.
\end{abstract}

\maketitle

%%%%%%%%%%%%%%%%%%%%%%%%%%%%%Main Body%%%%%%%%%%%%%%%%%%%%%%%%%%%%%%%%%%%%%

\section{Introduction \label{Sec:introduction}}

Superconductivity develops in metals as the result of Cooper pairing
instability when the attraction between electrons mediated by the
exchange of phonons (or other types of bosons) overcomes the static
Coulomb repulsion, which is the basic picture of
Bardeen-Cooper-Schrieffer (BCS) theory \cite{Schrieffer64}. In
principle, the precise values of the pairing gap $\Delta$ and the
transition temperature $T_c$ should be determined by performing a
careful theoretical study of the complicated interplay of
electron-phonon interaction (EPI) and Coulomb interaction. This is
difficult to achieve. Traditionally, these two interactions are
treated by using different methods \cite{Schrieffer64}. The EPI is
handled within the Migdal-Eliashberg (ME) theory, and $\Delta$ and
$T_c$ are computed by solving a set of integral equations satisfied
by the electrons' renormalization function and the pairing function.
In contrast, the Coulomb interaction is not handled at such a
quantitative level: its impact on $T_c$ is approximately measured by
one single pseudopotential parameter. Over the last decades, the ME
theory and the pseudopotential have been jointly applied
\cite{Scalapino, Allen, Carbotte, Marsiglio20} to evaluate $T_{c}$
and other quantities in various phonon-mediated superconductors.

That EPI and Coulomb interaction are handled quite differently can
be understood by making a field-theoretic analysis. Let us first
consider EPI. The EPI describes the mutual influence of the dynamics
of electrons and phonons on each other, and hence appears to be very
complicated. Within quantum many-body theory \cite{Schrieffer64,
Scalapino}, one needs to compute an infinite number of Feynman
diagrams to accurately compute any observable quantity, which is
apparently impractical. Fortunately, treatment of EPI can be greatly
simplified as the Migdal theorem \cite{Migdal} indicates that the
EPI vertex corrections are strongly suppressed by the small factor
$\lambda \left(\omega^{}_{\mathrm{D}}/E_{\mathrm{F}}\right)$, where
$\lambda$ is a dimensionless coupling parameter,
$\omega^{}_{\mathrm{D}}$ is Debye frequency, and $E_{\mathrm{F}}$ is
Fermi energy. For normal metals, $\lambda
\left(\omega^{}_{\mathrm{D}}/E_{\mathrm{F}}\right) \ll 1$, thus EPI
vertex corrections can be safely ignored. Under the bare vertex
approximation, Eliashberg \cite{Eliashberg} derived a set of coupled
equations, called ME equations, to study EPI-induced superconducting
transition.

We then discuss the influence of Coulomb repulsion. After defining
an auxiliary scalar field $A$ to represent the static Coulomb
potential, one can map the Coulomb interaction into an equivalent
fermion-boson interaction that has a similar field-theoretic
structure to EPI. But one cannot naively use the ME theory to handle
this fermion-boson interaction since there is no Migdal-like theorem
to guarantee the smallness of its vertex corrections. In the absence
of a well-controlled method, it seems necessary to make
approximations. Tolmachev \cite{Tolmachev} and Morel and Anderson
\cite{Morel62} introduced a pseudopotential $\mu^{\ast}$ to include
the impact of Coulomb repulsion. For a three-dimensional metal, the
bare Coulomb interaction is described by
\begin{eqnarray}
V_{0}(\mathbf{q}) = \frac{4\pi e^{2}}{\mathbf{q}^{2}}.
\end{eqnarray}
This bare function is renormalized to become energy-momentum
dependent, namely
\begin{eqnarray}
V_{R}(\omega,\mathbf{q}) = \frac{1}{V_{0}^{-1}(\mathbf{q}) -
\Pi(\omega,\mathbf{q})} = \frac{4\pi e^{2}}{\mathbf{q}^{2} - 4\pi
e^{2} \Pi(\omega,\mathbf{q})}.
\end{eqnarray}
The full polarization function $\Pi(\omega,\mathbf{q})$ is hard to
compute. A widely used approximation is to calculate
$\Pi(\omega,\mathbf{q})$ at the lowest one-loop level, corresponding
to the random phase approximation (RPA). The one-loop polarization
$\Pi_{1\mathrm{L}}(\omega,\mathbf{q})$ is still very complex.
Nevertheless, it is easy to reveal that
$\Pi_{1\mathrm{L}}(\omega,\mathbf{q})$ approaches a constant in the
limits of $\omega = 0$ and $\mathbf{q} \rightarrow 0$, i.e.,
$\Pi_{1\mathrm{L}}(\omega = 0,\mathbf{q} \rightarrow 0) \propto
-N_{0}$, where $N_{0}$ is the normal-state density of states (DOS)
on Fermi surface. For metals with a large Fermi surface, both
$E_{\mathrm{F}}$ and $N_{0}$ are fairly large. Thus the Coulomb
interaction becomes short-ranged and can be roughly described by
\cite{Morel62}
\begin{eqnarray}
V_{\mathrm{sim}}(\mathbf{q}) \propto \frac{1}{\mathbf{q}^{2} +
\kappa_{\mathrm{D}}^{}},\label{eq:vqrpa}
\end{eqnarray}
where the static screening factor $\kappa^{}_{\mathrm{D}} \propto
4\pi e^{2}N_{0}$. Morel and Anderson \cite{Morel62} suggested to
perform an average of $V_{\mathrm{sim}}(\mathbf{q})$ on the Fermi
surface, which yields a parameter $$\mu \propto \langle
V_{\mathrm{sim}}(\mathbf{q}) \rangle_{\mathrm{FS}}.$$ The
pseudopoential $\mu^{\ast}$ is related to $\mu$ via the relation
\cite{Morel62}
\begin{eqnarray}
\mu^{\ast}=\frac{\mu}{1+\mu
\ln(E_{\mathrm{F}}/\omega_{\mathrm{D}}^{})}.\label{eq:muast}
\end{eqnarray}
Obviously, $\mu^{\ast} \ll \mu$ in normal metals where
$\omega_{\mathrm{D}}^{} \ll E_{\mathrm{F}}$, rendering the
robustness of superconductivity against Coulomb repulsion.

As the above analysis demonstrates, the ME theory of EPI and the
Coulomb pseudopotential should be reliable if the condition
$\omega^{}_{\mathrm{D}} \ll E_{\mathrm{F}}$ is fulfilled. This
condition is violated in phonon-mediated superconductors that have a
low carrier density, with dilute SrTiO$_{3}$ being a famous example
\cite{Schooley64, Fernandes20}. In such superconductors,
$E_{\mathrm{F}}$ and $N_{0}$ are both small. There are no small
factors to suppress the EPI vertex corrections, indicating the
breakdown of Migdal theorem. Moreover, the Coulomb interaction is
poorly screened due to the smallness of $N_{0}$. The time-dependence
and spatial variation of Coulomb potential cannot be well described
by the oversimplified function $V_{\mathrm{sim}}(\mathbf{q})$ shown
in Eq.~(\ref{eq:vqrpa}). Accordingly, the pseugopotential defined in
Eq.~(\ref{eq:muast}) may no longer be valid as it comes directly
from Eq.~(\ref{eq:vqrpa}). It is more appropriate to adopt an
energy-momentum dependent $V_{R}(\omega,\mathbf{q})$ to replace the
static $V_{\mathrm{sim}}(\mathbf{q})$. Another potentially important
contribution arises from the mutual influence between EPI and
Coulomb interaction. This contribution was ignored in the original
work of Morel and Anderson \cite{Morel62} and also in most, if not
all, the subsequent studies on the Coulomb repulsion \cite{Sham83,
Richardson97, Katsnelson22, Prokefev22}. The validity of this
approximation is not clear. In principle, we expect that EPI can
affect Coulomb interaction and vice versa, because both EPI and
Coulomb interaction result in a re-distribution of all the charges
of the system. We should not simply discard their interplay if we
cannot prove that such an interplay is negligible. In light of the
above analysis, we consider it necessary to establish a more
powerful approach to supersede the ME theory of EPI and the
pseudopotential treatment of Coulomb repulsion. To achieve this
goal, we should take up the challenge of including all the higher
order corrections.

Recently, a non-perturbative Dyson-Schwinger (DS) equation approach
\cite{Liu21} was developed to determine the EPI vertex corrections.
At the core of this approach is the decoupling of the DS equation of
the full electron propagator $G(p)$ from all the rest DS equations
with the help of several exact identities. It is found \cite{Liu21}
that the DS equation of $G(p)$ obtained by using this approach is
self-closed and can be solved numerically. This approach was later
extended \cite{Pan21} to deal with one single fermion-boson
interaction, be it EPI or Coulomb interaction, in the context of
Dirac fermion systems. More recently, the approach was further
generalized \cite{Yang22} to investigate the coupling of Dirac
fermions to two different bosons. According to the results of
Refs.~\cite{Pan21, Yang22}, the DS equation of full Dirac fermion
propagator is self-closed irrespective of whether the fermions are
subjected to either EPI or Coulomb interaction, or both.

In this paper, we shall combine the approaches of \cite{Liu21} and
\cite{Yang22} to examine how the interplay of EPI and Coulomb
interaction affects the transition temperature $T_{c}$ of phonon
mediated superconductors. Our analysis will be based on an effective
model that describes the couplings of the electron field $\psi$ to a
phonon field $\phi$ and an auxiliary boson $A$. The EPI is described
by the $\psi$-$\phi$ coupling and the Coulomb interaction is
described by the $\psi$-$A$ coupling. Although there is not any
direct coupling between $\phi$ and $A$, these two bosons can affect
each other since they are both coupled to the same electrons. As a
consequence, the DS equation of $G(p)$ becomes formally very
complicated. To solve this difficulty, we derive four exact
identities after carrying out a series of analytical calculations
and then use such identities to show that the exact DS equation of
$G(p)$ is still self-closed. The higher order corrections neglected
in the ME theory and the pseudopotential method are properly taken
into account in this self-closed DS equation. Solving such an
equation leads us to $T_c$ and other quantities.

We shall apply the approach to a concrete example - the interfacial
superconductivity of one-unit-cell (1UC) FeSe/SrTiO$_{3}$ system.
After computing $T_{c}$ by solving the self-closed DS equation of
$G(p)$, we show that the value of $T_{c}$ depends strongly on the
chosen approximations. In particular, $T_{c}$ obtained under the
bare vertex (ME) approximation is substantially modified when the
vertex corrections are included.

The rest of the paper is organized as follows. In
Sec.~\ref{Sec:model}, we define the effective field theory for
phonon-mediated superconductors. In Sec.~\ref{Sec:DSE}, we obtain
the DS equation of $G(p)$ and prove its self-closure with the help
of four exact identities. In Sec.~\ref{Sec:numerics}, we present the
numerical results of $T_{c}$ obtained by solving the self-consistent
integral equations of two renormalization functions and the pairing
function. In Sec.~\ref{Sec:summary}, we summarize the results and
discuss the limitations of our calculations.

\section{Model \label{Sec:model}}

Our generic method is applicable to systems defined in any spatial
dimension. However, for concreteness, we consider a model defined at
two spatial dimensions, since later we shall apply the approach to
1UC FeSe/SrTiO$_{3}$. The interplay of EPI and Coulomb interaction
is described by the following effective Lagrangian density
\begin{eqnarray}
\mathcal{\mathcal{L}} &=& \mathcal{L}_{f}+\mathcal{L}_{p} +
\mathcal{L}_{A} + \mathcal{L}_{fp} + \mathcal{L}_{fA},
\label{eq:originallagrangian}\\
\mathcal{L}_{f} &=&  \psi^{\dag}(x)\left(i\partial_{x_{0}}\sigma_{0}
-\xi_{\nabla}\sigma_{3}\right)\psi(x), \\
\mathcal{L}_{p} &=& \frac{1}{2}\phi^\dag(x)\mathbb{D}(x)\phi(x),\\
\mathcal{L}_{A} &=& \frac{1}{2}A(x)\mathbb{F}(x)A(x),\\
\mathcal{L}_{fp} &=& - g\phi(x){\psi}^{\dag}(x)\sigma_{3}\psi(x),\\
\mathcal{L}_{fA} &=&- A(x){\psi}^{\dag}(x)\sigma_{3}\psi(x).
\end{eqnarray}
The electrons are represented by the Nambu spinor \cite{Nambu60}
$\psi({\mathbf p}) = \left(c_{\mathbf{p}\uparrow},
c_{-\mathbf{p}\downarrow}^{\dag}\right)^{T}$ along with four $2
\times 2$ matrices, including unit matrix $\sigma_{0}$ and three
Pauli matrices $\sigma_{1,2,3}$. Although the system is
non-relativistic, we choose to use a three-dimensional vector $x
\equiv (x_{0},\mathbf{x}) = (x_{0},x_{1},x_{2})$ for the purpose of
simplifying notations. The time $x_{0}$ can be either real or
imaginary (in Matsubara formalism), and the results hold in both
cases. The fermion field $\psi(x)$ is obtained by making Fourier
transformation to $\psi({\mathbf p})$. For simplicity, here we
assume that the kinetic energy operator is $\xi_{\nabla} =
-\frac{1}{2m_{e}}\left({\partial_{x_{1}}^{2}} +
{\partial_{x_{2}}^{2}}\right)-\mu_{\mathrm{F}}^{}$, where $m_{e}$ is the
bare electron mass and $\mu_{\mathrm{F}}^{}$ is the chemical
potential. As demonstrated in Ref.~\cite{Liu21}, our generic
approach remains valid if $\xi_{\nabla}$ takes a different form.
Phonons are represented by the scalar field $\phi(x)$, whose
equation of free motion is expressed via the operator
$\mathbb{D}(x)$ as
\begin{eqnarray}
\mathbb{D}(x)\phi(x)=0.
\end{eqnarray}
The EPI strength parameter $g$ appearing in $\mathcal{L}_{fA}$ is
not necessarily a constant and could be a function of phonon
momentum. $A(x)$ is an auxiliary scalar field. Its equation of free
motion is given by
\begin{eqnarray}
\mathbb{F}(x)A(x)=0.
\end{eqnarray}
The Coulomb interaction is effectively described by the coupling
between $\psi(x)$ and $A(x)$ shown in $\mathcal{L}_{fA}$. Indeed,
$\mathcal{L}_{A} + \mathcal{L}_{fA}$ can be derived by performing a
Hubbard-Stratonovich transformation to the following Hamiltonian
term for quartic Coulomb interaction
\begin{eqnarray}
\frac{e^{2}}{4\pi}\int d^2\mathbf{x} d^2 \mathbf{x}'{\psi^{\dag}}
(\mathbf{x}) \sigma_{3} \psi_{\sigma}(\mathbf{x})
\frac{1}{\left|\mathbf{x} -
\mathbf{x}'\right|}{\psi^{\dag}}(\mathbf{x}')\sigma_{3}
\psi(\mathbf{x}').
\end{eqnarray}

Notice that the model does not contain self-coupling terms of
bosons. The Coulomb interaction originates from the Abelian U(1)
gauge principle and the boson field $A(x)$ can be regarded as the
time component of U(1) gauge field (i.e., scalar potential). It is
well established that an Abelian gauge boson does not interact with
itself. The situation is different for phonons. In principle phonons
could interact with themselves. Ignoring the phonon self-couplings
is justified only when the lattice vibration is well captured by the
harmonic oscillating approximation. When the non-harmonic
contributions are not negligible, the self-couplings of phonons need
to be explicitly incorporated. Such non-harmonic contributions might
lead to a considerable influence on the value of $T_{c}$, as shown
in a recent work \cite{Johnstoncp}. In this paper, we shall not
consider the non-harmonic contributions and therefore omit
self-coupling terms of $\phi$.

Another notable feature of the model is that the two scalar fields
$\phi(x)$ and $A(x)$ do not directly interact with each other. Hence
there are no such terms as $\phi(x) A(x)$ or $\phi^{2}(x)A^{2}(x)$
in the Lagrangian density. However, the mutual influence between
$\phi(x)$ and $A(x)$ cannot be simply ignored since both of them are
coupled to the same electrons. It will be shown later that the DS
equation of electron propagator has a very complicated form due to
the mutual influence between $\phi(x)$ and $A(x)$. Moreover,
$\phi(x)$ and $A(x)$ are coupled to the same fermion density
operator $\psi^{\dag}(x)\sigma_{3}\psi(x)$. This implies that the
vertex function of $\psi$-$\phi$ coupling has a very similar
structure to that of $\psi$-$A$ coupling. The different behaviors of
$\phi$ boson and $A$ boson is primarily caused by the difference in
the operators $\mathbb{D}(x)$ and $\mathbb{F}(x)$, or equivalently,
the difference in the free propagators of $\phi$ boson and $A$
boson.

The Lagrangian density $\mathcal{L}$ respects the following two
global U(1) symmetries \cite{Nambu60}
\begin{eqnarray}
\psi &\rightarrow& e^{i\chi\sigma_3}\psi,\label{eq:csymmetry} \\
\psi &\rightarrow& e^{i\chi\sigma_0}\psi. \label{eq:ssymmetry}
\end{eqnarray}
Here, $\chi$ is an infinitesimal constant. The first symmetry
leads to charge conservation associated with a conserved current
$j^{c}_{\mu}(x) \equiv (j^{c}_{t}(x),\mathbf{j}^{c}(x))$, where
\begin{eqnarray}
j^{c}_{t}(x) &=& \psi^{\dag}(x)\sigma_{3}\psi(x), \\
\mathbf{j}^{c}(x) &=& \frac{1}{2m_{e}}\big[\left(i\mathbf{\nabla}
\psi^{\dag}(x)\right)\sigma_{0}\psi(x) - \psi^{\dag}(x)\sigma_{0}
\left(i\mathbf{\nabla}\psi(x)\right)\big].\nonumber \\
\label{eq:ccurrent}
\end{eqnarray}
The second symmetry leads to spin conservation and a conserved
current $j^{s}_{\mu}(x) \equiv (j^{s}_{t}(x),\mathbf{j}^{s}(x))$,
where
\begin{eqnarray}
j^{s}_{t}(x) &=& \psi^{\dag}(x)\sigma_{0}\psi(x), \\
\mathbf{j}^{s}(x) &=& \frac{1}{2m_{e}}\left[\left(i\mathbf{\nabla}
\psi^{\dag}(x)\right)\sigma_{3}\psi(x) - \psi^{\dag}(x)\sigma_{3}
\left(i\mathbf{\nabla}\psi(x)\right)\right].\nonumber \\
\label{eq:scurrent}
\end{eqnarray}
These two conserved currents obey the identity
$i\partial_{\mu}j^{c,s}_{\mu}(x)=0$ in the absence of external
sources. As shown in Ref.~\cite{Liu19}, each conserved current is
associated with one Ward-Takahashi identity (WTI).

\section{Dyson-Schwinger equation of electron propagator \label{Sec:DSE}}

After defining the effective model, we now are ready to perform a
non-perturbative study of the superconducting transition. The
following analysis will be largely based on the approaches
previously developed in Ref.~\cite{Liu21} and Ref.~\cite{Yang22}. We
shall not give all the derivational details and only outline the
major steps.

In order to examine the interaction-induced effects, we would like
to investigate the properties of various $n$-point correlation
functions. Such correlation functions can be generated from three
generating functionals \cite{Itzykson, Peskin}. Adding four external
sources $J$, $K$, $\eta$, and ${\eta^{{\dag}}}$ to the original
Lagrangian density leads to
\begin{eqnarray}
\mathcal{L}_{T} = \mathcal{L} + J\phi + K A +{\psi^{\dag}}\eta +
{\eta^{\dag}}\psi.
\end{eqnarray}
The partition function is defined via $\mathcal{L}_{T}$ as follows
\begin{eqnarray}
Z[J,K,\eta,{\eta^{\dag}}] \equiv \int D\phi DA D{\psi^{\dag}} D\psi
e^{i\int dx \mathcal{L}_{T}}.
\end{eqnarray}
Here, we use notation $\int dx$ to represent $\int d^{3}x = \int dt
d^{2}\mathbf{x}$. $Z$ is the generating functional for all $n$-point
correlation functions. In this paper, we are mainly interested in
connected correlation functions. Connected correlation functions can
be generated by the following generating functional
\begin{eqnarray}
W\equiv W[J,K,\eta,{\eta^{\dag}}] = -i\ln Z[J,K,\eta,{\eta^{\dag}}].
\end{eqnarray}
$W$ can be used to generate three two-point correlation functions
\begin{eqnarray}
G(x-y) = -i\langle\psi(x) {\psi^{\dag}}(y)\rangle = \frac{\delta^{2}
W}{\delta {\eta^{\dag}}(x)\delta \eta(y)}\Big|_{J=0}, \\
D(x-y) = -i\langle\phi(x)\phi^\dag(y)\rangle =
-\frac{\delta^{2}W}{\delta J(x)\delta J(y)}\Big|_{J=0}, \\
F(x-y) = -i\langle A(x)A(y)\rangle = -\frac{\delta^{2} W}{\delta
K(x) \delta K(y)}\Big|_{J=0}.
\end{eqnarray}
Here, $G(x-y)$, $D(x-y)$, and $F(x-y)$ are the full propagators of
electron $\psi$, phonon $\phi$, and boson $A$, respectively. The
system is supposed to be homogeneous, so the propagators depend
solely on the difference $x-y$. In this paper, we use the
abbreviated notation $J=0$ to indicate that all the external sources
are removed. All the correlation functions under consideration are
defined by the mean value of time-ordering product of various field
operators, but we omit the time-ordering symbols for simplicity. The
mutual influence between the properties of two bosons are embodied
in two additional two-point correlation functions
\begin{eqnarray}
D_{F}(x-y) = -i\langle\phi(x)A(y)\rangle =
-\frac{\delta^2W}{\delta J(x)\delta K(y)} \Big|_{J=0}, \\
F_{D}(x-y) = -i\langle A(x)\phi(y)\rangle = -\frac{\delta^2W}{\delta
K(x) \delta J(y)}\Big|_{J=0}.
\end{eqnarray}
As aforementioned, the model does not have such a term as $\phi A$,
thus $D_{F} = F_{D} = 0$ at the classic tree-level. But the quantum
(loop-level) corrections can induce non-zero contributions to
$D_{F}$ and $F_{D}$.

The interaction vertex function for a fermion-boson coupling can
also be generated from $W$. In the case of EPI, we consider the
following connected three-point correlation function:
%\begin{widetext}
\begin{eqnarray}
&& \langle\phi(x)\psi(y){\psi^{\dag}}(z)\rangle \nonumber \\
&=& \frac{\delta^3W}{\delta J(x)\delta \eta^{\dag}(y) \delta\eta(z)}
\Big|_{J=0} \label{eq:5} \nonumber \\
&=& -\int dx^{\prime} dy^{\prime} dz^{\prime} D(x-x^{\prime})
G(y-y^{\prime})\nonumber \\
&& \times \frac{\delta^3\Xi}{\delta\phi(x^{\prime})
\delta{\psi^{\dag}}(y^{\prime})\delta \psi(z^{\prime})}
\Big|_{J=0}G(z^{\prime}-z) \nonumber \\
&& -\int dx^{\prime} dy^{\prime} dz^{\prime} D_F(x-x^{\prime})
G(y-y^{\prime})\nonumber \\
&& \times \frac{\delta^3\Xi}{\delta A(x^{\prime})\delta
{\psi^{{\dag}}}(y^{\prime}) \delta\psi(z^{\prime})}\Big|_{J=0}
G(z^{\prime}-z),\label{eq:deltawdg}
\end{eqnarray}
%\end{widetext}
where the generating functional for proper (irreducible) vertices
$\Xi$ is defined via $W$ as
\begin{eqnarray}
\Xi = W - \int dx \Big[J\langle \phi\rangle + K\langle A\rangle +
\eta^{\dag}\langle\psi\rangle + \langle{\psi^{\dag}}
\rangle\eta\Big].
\end{eqnarray}
The interaction vertex function for EPI is defined as
\begin{eqnarray}
\Gamma_{p}(y-x,x-z) = \frac{\delta^3\Xi}{\delta\phi(x)
\delta{\psi^{{\dag}}}(y)\delta\psi(z)}\Big|_{J=0},
\end{eqnarray}
and that for $\psi$-$A$ coupling is defined as
\begin{eqnarray}
\Gamma_{A}(y-x,x-z) = \frac{\delta^3\Xi}{\delta A(x)
\delta{\psi^{{\dag}}}(y) \delta\psi(z)}\Big|_{J=0}.
\end{eqnarray}
It is necessary to emphasize that $\Gamma_{p}$ and $\Gamma_{A}$
depend on two (not three) free variables, namely $y-x$ and $x-z$.
The propagators and interaction vertex functions appearing in
Eq.~(\ref{eq:deltawdg}) are Fourier transformed as follows:
\begin{eqnarray}
G(p) &=& \int dx e^{ip x}G(x), \\
D(q) &=& \int dx e^{iq x}D(x), \\
D_F(q) &=& \int dx e^{iq x}D_F(x), \\
\Gamma_{p,A}(q,p) &=& \int dx dy e^{i(p+q)(y-x)}
e^{ip(x-z)} \nonumber \\
&& \times \Gamma_{p,A}(y-x,x-z),
\end{eqnarray}
Here, the electron momentum is $p \equiv (p_{0},\mathbf{p}) =
(p_{0},p_{1},p_{2})$ and the boson momentum is $q \equiv
(q_{0},\mathbf{q}) = (q_{0},q_{1},q_{2})$. Performing Fourier
transformation to $\langle\phi(x)\psi(y){\psi^{\dag}}(z)\rangle$, we
find
\begin{eqnarray}
&&\int dx dy e^{i(p+q)(y-x)}e^{ip(x-z)}
\langle\phi(x)\psi(y){\psi^{{\dag}}}(z)\rangle
\nonumber \\
&=& -D(q)G(p+q)\Gamma_p(q,p)G(p) \nonumber \\
&& -D_F(q)G(p+q)\Gamma_A(q,p)G(p). \label{eq:vertexphi}
\end{eqnarray}

The $\phi$-$A$ coupling can be investigated using the same
procedure. In this case, we need to study another three-point
correlation function $\langle A(x)\psi(y){\psi^{\dag}}(z)\rangle$.
Following the calculational steps that lead Eq.~(\ref{eq:deltawdg})
to Eq.~(\ref{eq:vertexphi}), we obtain
\begin{eqnarray}
&&\int dx dy e^{i(p+q)(y-x)}e^{ip(x-z)}
\langle A(x)\psi(y){\psi^{\dag}}(z)\rangle\nonumber\\
&=& -F(q)G(p+q)\Gamma_A(q,p)G(p) \nonumber \\
&& -F_D(q)G(p+q)\Gamma_p(q,p)G(p),
\end{eqnarray}
where $F(q)$ and $F_D(q)$ are transformed from $F(x)$ and $F_D(x)$
respectively as
\begin{eqnarray}
F(q) &=& \int dx e^{iq x}F(x), \\
F_{D}(q) &=& \int dx e^{iq x}F_{D}(x).
\end{eqnarray}

Making use of derivational procedure presented in Ref.~\cite{Liu21}
and Ref.~\cite{Yang22}, we find that the full electron propagator
$G(p)$ satisfies the following DS equation
\begin{eqnarray}
G^{-1}(p) &=& G_{0}^{-1}(p) - i\int \frac{d^{3}q}{(2\pi)^{3}}
g\sigma_{3} G(p+q)D(q)\Gamma_p(q,p) \nonumber
\\
&& -i\int \frac{d^{3}q}{(2\pi)^{3}} \sigma_{3}
G(p+q)F(q)\Gamma_A(q,p) \nonumber \\
&& -i \int \frac{d^{3}q}{(2\pi)^{3}} g\sigma_{3}
G(p+q)D_F(q)\Gamma_A(q,p) \nonumber \\
&& - i\int \frac{d^{3}q}{(2\pi)^{3}}
\sigma_{3}G(p+q)F_D(q)\Gamma_p(q,p).
\label{eq:dsegporiginal}
\end{eqnarray}
The electron self-energy $\Sigma(p) = G^{-1}(p) - G_{0}^{-1}(p)$
consists of four terms, as shown in the right-hand side (r.h.s.) of
Eq.~(\ref{eq:dsegporiginal}). The first two terms stem from pure EPI
and pure Coulomb interaction, respectively. The last two terms arise
from the mutual influence between these two interactions. The
contributions of such mixing terms to the self-energy were entirely
ignored in the original pseudopotential treatment of Morel and
Anderson \cite{Morel62}. To the best of our knowledge, such mixing
terms have never been seriously incorporated in previous
pseudopotential studies \cite{Sham83, Richardson97, Katsnelson22,
Prokefev22}. While ignoring them might be valid in some normal metal
superconductors, this approximation is not necessarily justified in
all cases. It would be better to keep them in calculations.

Unfortunately, retaining all the contributions to the self-energy
makes the DS equation of $G(p)$ extremely complex. It appears that
the equation (\ref{eq:dsegporiginal}) is not even self-closed since
$D(q)$, $F(q)$, $D_{F}(q)$, $F_{D}(q)$, $\Gamma_p(q,p)$, and
$\Gamma_{A}(q,p)$ are unknown. Technically, one can derive the DS
equations fulfilled by these six unknown functions by using the
generic rules of quantum field theory \cite{Itzykson, Peskin, Liu21,
Pan21, Yang22}. Nevertheless, such equations are coupled to the
formally more complicated DS equations of innumerable multi-point
correlation functions and hence of little use. Probably, one would
have to solve an infinite number of coupled DS equations to
completely determine $G(p)$, which is apparently not a feasible
scheme.

In order to simplify Eq.~(\ref{eq:dsegporiginal}) and make it
tractable, it might be necessary to introduce some approximations by
hands. For instance, one could: (1) neglect the last two (mixing)
terms of the r.h.s.; (2) discard all the vertex corrections by
assuming that $\Gamma_{p,A}(q,p)\rightarrow \sigma_{3}$; (3) replace
the full phonon propagator $D(q)$ with the bare one, i.e.,
$D(q)\rightarrow D_{0}(q)$; (4) replace the full $A$-boson
propagator $F(q)$ with a substantially simplified expression, such
as $F_{\mathrm{sim}}(q) = \frac{1}{\mathbf{q}^{2} +
\kappa_{\mathrm{D}}^{}}$, or even with one single (pseugopotential)
parameter $\mu^{\ast}$ after carrying out an average on the Fermi
surface. Under all of the above approximations, one find that the
original DS equation (\ref{eq:dsegporiginal}) becomes
\begin{eqnarray}
G^{-1}(p) &=& G_{0}^{-1}(p) - i\int \frac{d^{3}q}{(2\pi)^{3}}
g\sigma_{3} G(p+q)D_{0}(q)\sigma_{3} \nonumber \\
&& -i\mu^{\ast}\int \frac{d^{3}q}{(2\pi)^{3}} \sigma_{3}
G(p+q)\sigma_{3},\label{eq:megp}
\end{eqnarray}
which is self-closed and can be solved numerically. The free
electron propagator has the form
\begin{eqnarray}
G_{0}(p) = \frac{1}{i\epsilon_n\sigma_{0} -\xi_{\mathbf p}
\sigma_{3} },\label{eq:freegp}
\end{eqnarray}
and the full electron propagator is expanded as
\begin{eqnarray}
G(p) = \frac{1}{A_1(\epsilon_{n},\mathbf{p})i\epsilon_n\sigma_0 -
A_2(\epsilon_{n},\mathbf{p})\xi_{\mathbf p}\sigma_{3} +
\Delta(\epsilon_{n},\mathbf{p})\sigma_{1}},\label{eq:fullgp}
\nonumber \\
\end{eqnarray}
where $A_{1}(\epsilon_{n},\mathbf{p})$ and
$A_{2}(\epsilon_{n},\mathbf{p})$ are two renormalization functions
and $\Delta(\epsilon_{n},\mathbf{p})$ is pairing function. Inserting
$G(p)$ and $G_{0}(p)$ into Eq.~(\ref{eq:megp}), one would obtain the
standard ME equations of $A_{1}(\epsilon_{n},\mathbf{p})$,
$A_{2}(\epsilon_{n},\mathbf{p})$, and
$\Delta(\epsilon_{n},\mathbf{p})$ with the parameter $\mu^{\ast}$
characterizing the impact of Coulomb repulsion. In the past sixty
years, such simplified equations have been extensively applied
\cite{Schrieffer64, Scalapino, Allen, Carbotte, Marsiglio20} to
study a large number of phonon-mediated superconductors. However,
the four approximations that lead to Eq.~(\ref{eq:megp}) are not
always justified. Some, or perhaps all, of them break down in
superconductors having a small Fermi energy.

Now we seek to find a more powerful method to deal with the original
exact DS equation of $G(p)$ given by Eq.~(\ref{eq:dsegporiginal}) by
going beyond the above approximations. We believe that one should
not try to determine each of the six functions $D(q)$, $F(q)$,
$D_{F}(q)$, $F_{D}(q)$, $\Gamma_p(q,p)$, and $\Gamma_{A}(q,p)$
functions separately, which can never be achieved. Alternatively,
one should make an effort to determine such products as
$D(q)\Gamma_p(q,p)$, $F(q)\Gamma_A(q,p)$, $D_F(q)\Gamma_A(q,p)$, and
$F_D(q)\Gamma_p(q,p)$. This is the key idea of the approach proposed
in Ref.~\cite{Yang22}, where we have proved the self-closure of the
DS equation of the Dirac fermion propagator $G(p)$ in a model
describing the coupling of Dirac fermions to two distinct bosons in
graphene. Below we show that this same approach can be adopted to
prove the self-closure of the DS equation given by
Eq.~(\ref{eq:dsegporiginal}). We shall derive two exact identities
satisfied by $D(q)$, $F(q)$, $D_{F}(q)$, $F_{D}(q)$,
$\Gamma_p(q,p)$, and $\Gamma_{A}(q,p)$.

The derivation of the needed exact identities is based on the
invariance of partition function $Z$ under arbitrary infinitesimal
changes of $\phi$ and $A$. The invariance of $Z$ under an arbitrary
infinitesimal change of $\phi$ gives rise to
\begin{eqnarray}
\langle\mathbb{D}(x)\phi(x)-g{\psi^{\dag}}(x)\sigma_{3}\psi(x) +
J(x)\rangle = 0.\label{eq:eqmotionoriginal}
\end{eqnarray}
Using the relation $\langle \phi(x) \rangle = \delta W/\delta J(x)$,
we perform functional derivatives to the above equation with respect
to $\eta(z)$ and ${\eta^{{\dag}}}(y)$ in order and then obtain the
following new equation
\begin{eqnarray}
\mathbb{D}(x)\langle\phi(x)\psi(y){\psi^{{\dag}}}(z)\rangle =
g\langle {\psi^{{\dag}}}(x)\sigma_{3} \psi(x) \psi(y)
\psi^{{\dag}}(z)\rangle. \label{eq:W3anddensityvertex}
\nonumber \\
\end{eqnarray}
Making a Fourier transformation to the left-hand side (l.h.s.) of
Eq.~(\ref{eq:W3anddensityvertex}) yields
\begin{widetext}
\begin{eqnarray}
D_{0}^{-1}(q)\Big[-D(q)G(p+q)\Gamma_p(q,p)G(p) - D_{F}(q)G(p+q)
\Gamma_A(q,p)G(p)\Big],\label{eq:D0DGGammaG}
\end{eqnarray}
where the free phonon propagator $D_{0}(q)$ comes from
$\mathbb{D}(x)$. To handle the r.h.s. of
Eq.~(\ref{eq:W3anddensityvertex}), we use two bilinear operators
$j^{c}_{t}(x) = \psi^{\dag}(x)\sigma_{3}\psi(x)$ and $j^{s}_{t}(x) =
\psi^{\dag}(x)\sigma_{0}\psi(x)$ to define two current vertex
functions $\Gamma_{0,3}(x-z,z-y)$:
\begin{eqnarray}
\langle \psi^{\dag}(x)\sigma_{0,3}\psi(x)\psi(y)\psi^{\dag}(z)
\rangle = -\int d\zeta d\zeta' G(y-\zeta)
\Gamma_{0,3}(\zeta-x,x-\zeta') G(\zeta'-z). \label{eq:currentgp0}
\end{eqnarray}
$\Gamma_{0,3}(x-z,z-y)$ should be Fourier transformed
\cite{Engelsberg, Liu21} as
\begin{eqnarray}
\Gamma_{0,3}(\zeta-x,x-\zeta') = \int dq dp e^{-i(p+q) (\zeta-x)-ip
 (x-\zeta')}\Gamma_{0,3}(q,p). \label{eq:fouriergamma03}
\end{eqnarray}
For more properties of such current vertex functions, see
Refs.~\cite{Engelsberg, Liu21}. Then the r.h.s. of
Eq.~(\ref{eq:W3anddensityvertex}) is turned into
\begin{eqnarray}
\int dx dy e^{i(p+q)(y-x)} e^{ip(x-z)} g\langle
\psi^{{\dag}}(x)\sigma_{0,3} \psi(x)\psi(y)\psi^{{\dag}}(z) \rangle
\rightarrow -g G(p+q)\Gamma_{0,3}(q,p)G(p). \label{eq:gGGammaG}
\end{eqnarray}
After substituting Eq.~(\ref{eq:D0DGGammaG}) and
Eq.~(\ref{eq:gGGammaG}) into Eq.~(\ref{eq:W3anddensityvertex}), we
obtain the following identity
\begin{eqnarray}
D(q)\Gamma_p(q,p)+D_F(q)\Gamma_A(q,p) = D_0(q)g\Gamma_{3}(q,p).
\label{eq:d0gamma3}
\end{eqnarray}
Similarly, the invariance of $Z$ under an infinitesimal change of
$A$ field requires the following equation to hold
\begin{eqnarray}
\langle\mathbb{F}(x)A(x)-g{\psi^{\dag}}(x)\sigma_{3}\psi(x) +
K(x)\rangle = 0.
\end{eqnarray}
Carrying out similar analytical calculations generates another
identity
\begin{eqnarray}
F_D(q)\Gamma_p(q,p)+F(q)\Gamma_A(q,p) = F_0(q)\Gamma_{3}(q,p),
\label{eq:f0gamma3}
\end{eqnarray}
where the free propagator of $A$ boson $F_0(q)$ is computed by
performing Fourier transformation to $\mathbb{F}(x)$.

Making use of the two identities given by Eq.~(\ref{eq:d0gamma3})
and Eq.~(\ref{eq:f0gamma3}), we re-write the original DS equation
(\ref{eq:dsegporiginal}) as
\begin{eqnarray}
G^{-1}(p) = G_{0}^{-1}(p) - i\int \frac{d^{3}q}{(2\pi)^{3}}
\left[g^2 D_0(q) + F_0(q)\right] \sigma_{3} G(p+q)
\Gamma_{3}(q,p).\label{eq:dsefinal}
\end{eqnarray}
This equation is still not self-closed if the current vertex
function $\Gamma_{3}(q,p)$ relies on unknown functions other than
$G(p)$. As demonstrated in the Supplementary Material
\cite{supplementary} (see also references \cite{Dirac, Schwinger,
Jackiw69, Bardeen69, Callan70, Schnabl, He01} therein), the symmetry
of Eq.~(\ref{eq:csymmetry}) leads to the following WTI
\begin{eqnarray}
q_{0}\Gamma_{3}(q,p)-\left(\xi_{\mathbf{p+q}}-\xi_{\mathbf{p}}\right)
\Gamma_{0}(q,p) &=& G^{-1}(p+q)\sigma_{3}-\sigma_{3}G^{-1}(p).
\label{eq:chargewti}
\end{eqnarray}
It is not possible to determine $\Gamma_{3}(q,p)$ purely based on
this single WTI, since $\Gamma_{0}(q,p)$ is also unknown.
Fortunately, from \cite{supplementary} (see also references
\cite{Dirac, Schwinger, Jackiw69, Bardeen69, Callan70, Schnabl,
He01} therein) we know that the symmetry of Eq.~(\ref{eq:ssymmetry})
yields another WTI
\begin{eqnarray}
q_{0}\Gamma_{0}(q,p)-\left(\xi_{\mathbf{p+q}}-\xi_{\mathbf{p}}\right)
\Gamma_{3}(q,p) &=& G^{-1}(p+q)\sigma_{0} - \sigma_{0}G^{-1}(p).
\label{eq:spinwti}
\end{eqnarray}
These two WTIs are coupled to each other and can be used to express
$\Gamma_{3}(q,p)$ and $\Gamma_{0}(q,p)$ purely in terms of $G(p)$.
Now $\Gamma_{3}(q,p)$ can be readily obtained by solving these two
WTIs, and its expression is
\begin{eqnarray}
\Gamma_{3}(q,p) = \frac{q_{0}\left[G^{-1}(p+q)\sigma_3 -\sigma_3
G^{-1}(p)\right]+\left(\xi_{\mathbf{p}+\mathbf{q}} -
\xi_{\mathbf{p}}\right)\left[G^{-1}(p+q)\sigma_0 - \sigma_0
G^{-1}(p)\right]}{q_{0}^{2}-\left(\xi_{\mathbf{p}+\mathbf{q}} -
\xi_{\mathbf{p}}\right)^2}. \label{eq:gamma3qp}
\end{eqnarray}
\end{widetext}

We can see that the DS equation (\ref{eq:dsefinal}) becomes entirely
self-closed because it contains merely one unknown function $G(p)$.
This equation can be numerically solved to determine $G(p)$,
provided that $G_{0}(p)$, $D_{0}(q)$, and $F_{0}(q)$, and $g$ are
known.

It is useful to make some remarks here:

(1) The two WTIs given by Eq.~(\ref{eq:chargewti}) and
Eq.~(\ref{eq:spinwti}) were originally obtained in Ref.~\cite{Liu21}
based on a pure EPI model. The model considered in this work
contains an additional fermion-boson coupling that equivalently
represents the Coulomb interaction. We emphasize that such an
addition coupling does not alter the WTIs, since the Lagrangian
density of pure EPI and the one describing the interplay between EPI
and Coulomb interaction preserve the same U(1) symmetries defined by
Eq.~(\ref{eq:csymmetry}) and Eq.~(\ref{eq:ssymmetry}).

(2) In many existing publications, it is naively deemed that a
symmetry-induced WTI imposes an exact relation between fermion
propagator $G(p)$ and interaction vertex function. To understand why
this is a misconception, let us take EPI as an example. The EPI
vertex function $\Gamma_{p}(q,p)$ is defined via a three-point
correlation function $\langle \phi\psi\psi^{\dag}\rangle$, which in
itself is not necessarily related to any conserved current. There is
no reason to expect $\Gamma_{p}(q,p)$ to naturally enter into any
WTI. To reveal the impact of some symmetry, one should use the
symmetry-induced conserved current, say $j_{\mu}^{c}$, to define a
special correlation function $\langle
j_{\mu}^{c}\psi\psi^{\dag}\rangle$, which, according to
Eq.~(\ref{eq:currentgp0}) and Eq.~(\ref{eq:gGGammaG}), is expressed
in terms of current vertex functions $\Gamma_{0}(q,p)$ and
$\Gamma_{3}(q,p)$. After applying the constraint of current
conservation $\partial_{\mu} j_{\mu}^{c} = 0$ to $\langle
j_{\mu}^{c}\psi\psi^{\dag}\rangle$, one would obtain a WTI satisfied
by $\Gamma_{0}(q,p)$, $\Gamma_{3}(q,p)$, and $G(p)$, as shown in
Eq.~(\ref{eq:chargewti}).

(3) Our ultimate goal is to determine $G(p)$. Its DS equation
(\ref{eq:dsegporiginal}) contains two interaction vertex functions
$\Gamma_{p}(q,p)$ and $\Gamma_{A}(q,p)$. On the other hand, it is
$\Gamma_{0}(q,p)$ and $\Gamma_{3}(q,p)$, rather than
$\Gamma_{p}(q,p)$ and $\Gamma_{A}(q,p)$, that enter into the WTIs
given by Eq.~(\ref{eq:chargewti}) and Eq.~(\ref{eq:spinwti}). Hence,
at least superficially the DS equation of $G(p)$ and the WTIs are
not evidently correlated. To find out a natural way to combine the
DS equation of $G(p)$ with WTIs, one needs to obtain the relations
between interaction vertex functions and current vertex functions.
Such relations do exist and are shown in Eq.~(\ref{eq:d0gamma3}) and
Eq.~(\ref{eq:f0gamma3}).

(4) The appearance of two free boson propagators $D_{0}(q)$ and
$F_{0}(q)$ in the final DS equation (\ref{eq:dsefinal}) is not an
approximation. It should be emphasized that the replacement of the
full boson propagators $D(q)$ and $F(q)$ appearing in the original
DS equation (\ref{eq:dsegporiginal}) with their free ones is
implemented based on two exact identities given by
Eq.~(\ref{eq:d0gamma3}) and Eq.~(\ref{eq:f0gamma3}). The
interaction-induced effects embodied in such functions as $D(q)$,
$F(q)$, $D_{F}(q)$, $F_{D}(q)$, $\Gamma_{p}(q,p)$, and
$\Gamma_{A}(q,p)$ are already incorporated in current vertex
function $\Gamma_{3}(q,p)$.

Before closing this section, we briefly discuss whether our approach
is applicable to four-fermion interactions. The Hubbard model is a
typical example of this type. Consider a simple four-fermion
coupling term given by
\begin{eqnarray}
U_{H}\Psi^{\dag}\Psi\Psi^{\dag}\Psi,\label{eq:hubbardmodel}
\end{eqnarray}
where $\Psi$ is a normal (non-Nambu) spinor. Based on this Hubbard
model, one can derive DS integral equations and WTIs satisfied by
various correlation functions. Actually, it is straightforward to
obtain a U(1)-symmetry-induced WTI that connects the fermion
propagator $G(p)$ to a current vertex function $\Gamma_{H}(p,p+q)$
defined through the following correlation function
\begin{eqnarray}
\langle j_{\mu} \Psi\Psi^{\dag}\rangle \rightarrow
G(p_{1})\Gamma_{H}(p_{1},p_{2})G(p_{2}),
\end{eqnarray}
where $j_{\mu}$ is a conserved (charge) current operator. The
fermion propagator $G(p)$ should be determined by solving its DS
integral equation. As demonstrated in Ref.~\cite{AGDbook}, the DS
equation of $G(p)$ contains a two-particle kernel function
$\Gamma_{4}(p_{1},p_{2},p_{3},p_{4})$, which is defined via a
four-point correlation function as follows
\begin{eqnarray}
\langle \Psi^{\dag}\Psi\Psi^{\dag}\Psi \rangle \rightarrow
G(p_{1})G(p_{2})\Gamma_{4}(p_{1},p_{2},p_{3},p_{4})G(p_{3})G(p_{4}).
\nonumber
\end{eqnarray}
It is clear that $\Gamma_{H}$ is physically distinct from
$\Gamma_{4}$. We are not aware of the presence of any simple
relation between these two functions. A field-theoretic analysis
reveals that the DS integral equation of $\Gamma_{4}$ is strongly
coupled to an infinite number of DS integral equations of various
$n$-point correlation functions ($n > 4$). Even though $\Gamma_{H}$
can be expressed purely in terms of $G(p)$ after solving a number of
WTIs, it cannot be used to simplify the DS equation of $G(p)$
because of our ignorance of the structure of $\Gamma_{4}$. It is
therefore unlikely that our approach is directly applicable to
Hubbard-type models like Eq.~(\ref{eq:hubbardmodel}).

Alternatively, one could introduce an auxiliary bosonic field
$\varphi$ and then perform a Hubbard-Strachnovich transformation,
which turns the original Hubbard model into a Yukawa-type
fermion-boson coupling term
\begin{eqnarray}
\mathcal{L}_{\mathrm{Y}} = -g_{Y}\varphi
\Psi^{\dag}\Psi.\label{eq:effectiveyukawacoupling}
\end{eqnarray}
It seems that this coupling could be treated in the same way as what
we have done for the Coulomb interaction. However, we emphasize that
this Yukawa coupling alone cannot describe all the physical effects
produced by the original Hubbard four-fermion coupling. This is
because boson self-interactions cannot be simply neglected. In the
case of Coulomb interaction, the Abelian U(1) gauge invariance
guarantees the absence of self-interactions of $a_{0}$ bosons. In
contrast, there is not any physical principle to prevent the
auxiliary boson field $\varphi$ from developing such a self-coupling
term:
\begin{eqnarray}
\mathcal{L}_{4} = u_{4} \varphi^{4}(x).\label{eq:varphi4coupling}
\end{eqnarray}
In quantum field theory, it is well-established \cite{Peskin} that
the Yukawa interaction cannot be renormalized if the model does not
contain an appropriate quartic term. In condensed matter physics,
the boson self-interactions have been found \cite{Chubukov04,
Metlitski10, Liu19, Torroba20} to play a significant role,
especially in the vicinity of a quantum critical point. In fact,
even if the Lagrangian originally does not contain any boson
self-coupling term, the Yukawa coupling $g_{Y}\varphi
\Psi^{\dag}\Psi$ can dynamically generating certain boson
self-coupling terms \cite{Chubukov04, Liu19}. After including the
term $\sim \varphi^{4}$, the invariance of $Z$ under an arbitrary
infinitesimal change of $\varphi$ leads to
\begin{eqnarray}
\langle\mathbb{D}(x)\varphi(x)+4u_{4}\varphi^{3}(x) -
g_{Y}{\Psi^{\dag}}(x) \Psi(x) + J(x)\rangle = 0.
\label{eq:eqmotionquartic}
\end{eqnarray}
Comparing to Eq.~(\ref{eq:eqmotionoriginal}), there appears an
additional term $\sim u_{4}\varphi^{3}(x)$ owing to the boson
self-interaction. After performing functional derivatives with
respect to $\eta(z)$ and ${\eta^{{\dag}}}(y)$ in order, one obtains
\begin{eqnarray}
&&\mathbb{D}(x)\langle\varphi(x)\Psi(y){\Psi^{{\dag}}}(z)\rangle +
4u_{4}\langle\varphi^{3}(x)\Psi(y){\Psi^{{\dag}}}(z)\rangle \nonumber \\
&=& g_{Y}\langle {\Psi^{{\dag}}}(x)\Psi(x) \Psi(y)
\Psi^{{\dag}}(z)\rangle. \label{eq:W3anddensityvertexquartic}
\end{eqnarray}
Different from Eq.~(\ref{eq:W3anddensityvertex}), this equation
contains an extra correlation function $\langle \varphi^{3}(x)
\Psi(y) {\Psi^{{\dag}}}(z)\rangle$. This correlation function is
formally very complicated and actually makes it impossible to derive
a self-closed DS equation of the fermion propagator. Thus, our
approach is applicable only when the quartic term $\sim \varphi^{4}$
can be safely ignored.

\section{Numerical results of $T_{c}$ \label{Sec:numerics}}

In this section, we apply the self-closed DS equation of $G(p)$
given by Eq.~(\ref{eq:dsefinal}) along with Eq.~(\ref{eq:gamma3qp})
to evaluate the pair-breaking temperature $T_{c}$ of the
superconductivity realized in 1UC FeSe/SrTiO$_{3}$ system. This
material is found \cite{Xue12, Shen14, Lee15} to possess a
surprisingly high $T_{c}$. While it is believed by many
\cite{Shen14, Lee15} that interfacial optical phonons (IOPs) from
the SrTiO$_{3}$ substrate are responsible for the observed high
$T_{c}$, other microscopic pairing mechanisms cannot be conclusively
excluded. Gor'kov \cite{Gorkov16} argued that IOPs alone are not
capable of causing such a high $T_{c}$. If this conclusion (not
necessarily the argument itself) is reliable, we would be compelled
to invoke at least one additional pairing mechanism, such as
magnetic fluctuation or nematic fluctuation, to cooperate with IOPs
\cite{Lee15, Gorkov16, Yao16}. In recent years, considerable
research efforts have been devoted to calculating IOPs-induced
$T_{c}$ by using the standard ME theory \cite{Xiang12, Xing14,
Johnston-NJP2016, Dolgov17, Opp-PRB2018} and a slightly corrected
version of ME theory \cite{Opp-PRB2021}. Nevertheless, thus far no
consensus has been reached and the accurate value of $T_{c}$
produced by IOPs alone is still controversial. To get a definite
answer, it is important to compute $T_{c}$ with a sufficiently high
precision. This is certainly not an easy task since $T_{c}$ could be
influenced by many factors.

Among all the factors that can potentially affect $T_{c}$, the EPI
vertex corrections play a major role. Since the ratio
$\omega^{}_{\mathrm{D}}/E_{\mathrm{F}}$ is at the order of unity,
the Migdal theorem becomes invalid. As shown in Ref.~\cite{Liu21},
including EPI vertex corrections can drastically change the value of
$T_{c}$ obtained under bare vertex approximation. However, the
calculations of Ref.~\cite{Liu21} were based on two approximations
that might lead to inaccuracies and thus still need to be improved.
The first approximation is the omission of the influence of Coulomb
repulsion \cite{Liu21}. As discussed in Sec.~\ref{Sec:introduction},
the traditional pseudopotential method may not work well in 1UC
FeSe/SrTiO$_{3}$ owing to the smallness of $E_{\mathrm{F}}$. The
impact of Coulomb interaction on $T_{c}$ should be examined more
carefully. The second approximation is that the electron momentum
was supposed \cite{Liu21} to be fixed at the Fermi momentum such
that $\xi_{\mathbf{p}}=0$. Under this second approximation, the DS
equation of $G(p)$ has only one integral variable (i.e., frequency).
Then the computational time is significantly shortened. For this
reason, this kind of approximation has widely been used in the
existing calculations of $T_{c}$. But the pairing gap $\Delta$ and
the renormalization factors $A_{1}$ and $A_{2}$ obtained by solving
their single-variable equations depend solely on frequency. The
momentum dependence is entirely lost. Since the EPI strength depends
strongly on the transferred momentum $\mathbf{q}$, it is important
not to neglect the momentum dependence of the DS equation of $G(p)$.
We shall discard the two approximations adopted in Ref.~\cite{Liu21}
and directly deal with the self-closed DS equation
(\ref{eq:dsefinal}).

We are particularly interested in how $T_{c}$ is affected by the
interplay of EPI and Coulomb repulsion. To avoid the difficulty
brought by analytical continuation, we work in the Matsubara
formalism and express the electron momentum as
$p\equiv(p_{0},\mathbf{p}) = (i\epsilon_n,\mathbf{p})$, where
$\epsilon_n = (2n+1)\pi T$, and the boson momentum as
$q\equiv(q_{0},\mathbf{q}) = (i\omega_{n^{\prime}},\mathbf{q})$,
where $\omega_{n^{\prime}} = 2n^{\prime}\pi T$. The free phonon
propagator has the form
\begin{eqnarray}
D_{0}(q) = \frac{2\Omega_{\mathbf{q}}}{(i\omega_{n'})^2 -
\Omega_{\mathbf{q}}^2}.\label{eq:freedq}
\end{eqnarray}
The IOPs are found to be almost dispersionless \cite{Shen14,
Gongxg15}, thus $\Omega_{\mathbf{q}}$ can be well approximated by a
constant. Here, we choose the value \cite{Shen14, Gongxg15}
$\Omega_{\mathbf{q}} = 81 \mathrm{~meV}$. The Fermi energy is roughly
\cite{Lee15} is $E_{\mathrm{F}}=65\mathrm{~meV}$. The EPI strength
parameter $g$ is related to phonon momentum $\mathbf{q}$ as
\cite{Johnston-NJP2016}
\begin{eqnarray}
g \equiv g(\mathbf{q}) = \sqrt{8\pi\lambda/q_{\mathrm{p}}^2}
\Omega_{\mathbf{q}}\exp(-|\mathbf{q}|/q_{\mathrm{p}}).
\label{eq:gq}
\end{eqnarray}
The value of $\lambda$ can be estimated by first-principle
calculations \cite{Johnston-NJP2016}. Here we regard $\lambda$ as a
tuning parameter and choose a set of different values in our
calculations. The range of EPI is characterized by the parameter
$q_{\mathrm{p}}$ \cite{Lee15}. Its precise value relies on the
values of other parameters and is hard to determine. For simplicity,
we choose to fix its value \cite{Johnston-NJP2016} at
$q_{\mathrm{p}} = 0.1 ~p_{\mathrm{F}}$. The free propagator of
$A$-boson is
\begin{eqnarray}
F_{0}(q) = \frac{2\pi \alpha}{|\mathbf{q}|},\label{eq:freefq}
\end{eqnarray}
which has the same form as the bare Coulomb interaction function.
The fine structure constant is $\alpha = e^{2}/v_{F} \varepsilon$.
The magnitude of dielectric constant $\varepsilon$ depends
sensitively on the surroundings (substrate) of superconducting film.
It is not easy to accurately determine $\varepsilon$. To make our
analysis more general, we suppose that $\varepsilon$ can be freely
changed within a certain range.

As the next step, we wish to substitute the free phonon propagator
$D_{0}(q)$ given by Eq.~(\ref{eq:freedq}), the free $A$-boson
propagator $F_{0}(q)$ given by Eq.~(\ref{eq:freefq}), the free
electron propagator $G_{0}(p)$ given by Eq.~(\ref{eq:freegp}), and
the full electron propagator $G(p)$ given by Eq.~(\ref{eq:fullgp})
into the DS equation (\ref{eq:dsefinal}) and also into the function
$\Gamma_{3}(q,p)$ given by Eq.~(\ref{eq:gamma3qp}). However, we
cannot naively do so since here we encounter a fundamental problem.
Recall that $\Gamma_{3}(q,p)$ given by Eq.~(\ref{eq:gamma3qp}) is
derived from two symmetry-induced WTIs. Once the pairing function
$\Delta(p)$ develops a finite value, the system enters into
superconducting state. The U(1) symmetry of Eq.~(\ref{eq:ssymmetry})
is preserved in both the normal and superconducting states, thus the
WTI of Eq.~(\ref{eq:spinwti}) is not changed. In contrast, the U(1)
symmetry of Eq.~(\ref{eq:csymmetry}) is spontaneously broken in the
superconducting state. If this symmetry breaking does not change the
WTI of Eq.~(\ref{eq:chargewti}), one could insert the expression of
$G(p)$ given by Eq.~(\ref{eq:fullgp}) into $\Gamma_{3}(q,p)$.
Otherwise, one should explore the modification of the WTI by
symmetry breaking. At present, there seems no conclusive answer.
Nambu \cite{Nambu60} adopted the WTI from charge conservation to
prove the gauge invariance of electromagnetic response functions of
a superconductor based on a ladder-approximation of the DS equation
of vertex function. Following the scheme of Nambu, Schrieffer
\cite{Schrieffer64} assumed (without giving a proof) that this WTI
is the same in the superconducting and normal phases and used this
assumption to show the existence of a gapless Goldstone mode.
Nakanishi \cite{Nakanishi} later demonstrated that, while the WTI
for a U(1) gauge field theory has the same form in symmetric and
symmetry-broken phases, it might be altered in other field theories.
Recently, Yanagisawa \cite{Yanagisawa} revisited this issue and
argued that the spontaneous breaking of a continuous symmetry gives
rise to an additional term to WTI due to the generation of Goldstone
boson(s). However, the expression of such an additional term is
unknown. It also remains unclear whether the approach of
Ref.~\cite{Yanagisawa} still works in superconductors. The
superconducting transition is profoundly different from other
symmetry-breaking driven transitions. According to the Anderson
mechanism \cite{Anderson63}, the Goldstone boson generated by
U(1)-symmetry breaking is eaten by the long-range Coulomb
interaction, which lifts the originally gapless mode to a gapped
plasmon mode. Thus, the WTI coming from symmetry
Eq.~(\ref{eq:csymmetry}) is not expected to acquire the additional
term derived in Ref.~\cite{Yanagisawa} in the superconducting phase.
Nevertheless, the absence of Goldstone-boson cannot ensure that the
WTI is not changed by Anderson mechanism.

In order to attain a complete theoretical description of
superconducting transition, one should strive to develop a unified
framework to reconcile the non-perturbative DS equation approach
with the Anderson mechanism. But such a framework is currently not
available. To proceed with our calculations, we have to introduce a
suitable approximation. Our purpose is to compute $T_{c}$. Near
$T_{c}$, the pairing function $\Delta(p)$ vanishes and the symmetry
Eq.~(\ref{eq:csymmetry}) is still preserved. So the WTI of
Eq.~(\ref{eq:chargewti}) still holds. Then we substitute the full
electron propagator $G(p)$ given by Eq.~(\ref{eq:fullgp}) into
Eq.~(\ref{eq:dsefinal}) and assume the function $\Gamma_{3}(q,p)$ to
have the following expression
\begin{widetext}
\begin{eqnarray}
\Gamma_{3}(q,p) = \frac{q_{0}\left[G_{s}^{-1}(p+q)\sigma_3 -\sigma_3
G_{s}^{-1}(p)\right]+\left(\xi_{\mathbf{p}+\mathbf{q}} -
\xi_{\mathbf{p}}\right)\left[G_{s}^{-1}(p+q)\sigma_0 - \sigma_0
G_{s}^{-1}(p)\right]}{q_{0}^{2}-\left(\xi_{\mathbf{p}+\mathbf{q}} -
\xi_{\mathbf{p}}\right)^2}, \label{eq:gamma3qpsimplified}
\end{eqnarray}
where $G_{s}(p)$ is a simplified electron propagator of the form
\begin{eqnarray}
G_{s}(p) = \frac{1}{A_1(p)p_{0}\sigma_0 - A_2(p)\xi_{\mathbf p}
\sigma_{3}}.
\end{eqnarray}
This manipulation leads to three coupled nonlinear integral
equations:
\begin{eqnarray}
A_1(\epsilon_n,{\mathbf p}) &=& 1 + \frac{T}{i\epsilon_n}\sum_{m}
\int \frac{d^2 {\mathbf q}}{(2\pi )^2}\Big[g^2({\mathbf
q})D_0(\omega_m,{\mathbf q}) + F_0(\omega_m,{\mathbf
q})\Big]\nonumber \\
&&\times \frac{A_1(\epsilon_n+\omega_m,{\mathbf{p+q}}) i(\epsilon_n
+ \omega_m)\Gamma_{33} + A_2(\epsilon_n+\omega_m,{\mathbf{p+q}})
\xi_{\mathbf{p+q}}\Gamma_{30} - \Delta(\epsilon_n+\omega_m,
{\mathbf{p+q}}) \Gamma_{3d}}{A_1^2(\epsilon_n+\omega_m,
{\mathbf{p+q}})(\epsilon_n+\omega_m)^2 + A_2^2(\epsilon_n+\omega_m,
{\mathbf{p+q}})\xi^2_{\mathbf{p+q}} +
\Delta^2(\epsilon_n+\omega_m,\mathbf{p+q})}, \label{eq:A1p}\\
A_2(\epsilon_n,{\mathbf p}) &=& 1 - \frac{T}{\xi_{\mathbf{p}}}
\sum_{m}\int \frac{d^2{\mathbf q}}{(2\pi)^2}\Big[g^2({\mathbf q})
D_0(\omega_m,{\mathbf q})+F_0(\omega_m,{\mathbf q})\Big]
\nonumber \\
&&\times \frac{A_1(\epsilon_n+\omega_m,{\mathbf{p+q}})
i(\epsilon_n+\omega_m)\Gamma_{30}+A_2(
\epsilon_n+\omega_m,{\mathbf{p+q}})\xi_{\mathbf{p+q}}\Gamma_{33} -
\Delta({\epsilon_n+\omega_m,\mathbf{p+q}})
\Gamma_{31}}{A_1^2(\epsilon_n+\omega_m,{\mathbf{p+q}})
(\epsilon_n+\omega_m)^2+A_2^2(\epsilon_n + \omega_m, {\mathbf{p+q}})
\xi^2_{\mathbf{p+q}} + \Delta^2
(\epsilon_n+\omega_m,{\mathbf{p+q}})}, \label{eq:A2p} \\
\Delta(\epsilon_n,{\mathbf p}) &=& -T \sum_{m}\int \frac{d^2{\mathbf
q}}{(2\pi)^2} \Big[g^2({\mathbf q})D_0(\omega_m,{\mathbf q}) +
F_0(\omega_m,{\mathbf q})\Big] \nonumber \\
&&\times \frac{A_1(\epsilon_n+\omega_m,{\mathbf{p+q}})
i(\epsilon_n+\omega_m)\Gamma_{3d} - A_2( \epsilon_n +
\omega_m,{\mathbf{p+q}})\xi_{\mathbf{p+q}}\Gamma_{31} -
\Delta(\epsilon_{n}+\omega_{m},{\mathbf{p+q}})\Gamma_{33}}
{A_{1}^{2}(\epsilon_{n}+\omega_m,\mathbf{p+q})(\epsilon_n +
\omega_{m})^{2} + A_2^2(\epsilon_n+\omega_m,{\mathbf{p+q}})
\xi^2_{\mathbf{p+q}}+\Delta^2(\epsilon_n+\omega_m,{\mathbf{p+q}})}.
\label{eq:Deltap}
\end{eqnarray}
Here, we have defined several quantities:
\begin{eqnarray}
\Gamma_{30} &=& \frac{i\omega_m \left[A_2(\epsilon_n+\omega_m,
{\mathbf{p+q}})\xi_{\mathbf{p+q}} - A_2(\epsilon_n,{\mathbf p})
\xi_{\mathbf p}\right] + (\xi_{\mathbf{p+q}}-\xi_{\mathbf p})
\left[-A_1(\epsilon_n + \omega_m,{\mathbf{p+q}})
(i\epsilon_n+i\omega_m) + A_1(\epsilon_n, {\mathbf p})
i\epsilon_n\right]}{\omega_m^2+(\xi_{\mathbf{p+q}}
- \xi_{\mathbf p})^2},\label{eq:gamma30} \nonumber \\
\Gamma_{33} &=& \frac{i\omega_m\left[-A_1(\epsilon_n +
\omega_m,{\mathbf{p+q}})(i\epsilon_n+i\omega_m) + A_1(\epsilon_n,
{\mathbf p})i\epsilon_n\right]+(\xi_{\mathbf{p+q}}-\xi_{\mathbf p})
\left[A_2 (\epsilon_n+\omega_m,{\mathbf{p+q}})\xi_{\mathbf{p+q}} -
A_2(\epsilon_n,{\mathbf p})\xi_{\mathbf p}\right]}{\omega_m^2 +
(\xi_{\mathbf{p+q}} - \xi_{\mathbf p})^2}, \label{eq:gamma33}
\nonumber \\
\Gamma_{31} &=& \frac{(\xi_{\mathbf{p+q}}-\xi_{\mathbf p})
\left[-\Delta(\epsilon_n+\omega_m,{\mathbf{p+q}})+\Delta(\epsilon_n,
{\mathbf p})\right]}{\omega_{m}^{2}+(\xi_{\mathbf{p+q}} -
\xi_{\mathbf p})^2}, \label{eq:gamma31} \nonumber \\
\Gamma_{3d} &=& \frac{i\omega_{m}\left[\Delta(\epsilon_{n} +
\omega_m,{\mathbf{p+q}})+\Delta(\epsilon_n,{\mathbf p})\right]}
{\omega_{m}^{2}+(\xi_{\mathbf{p+q}}-\xi_{\mathbf p})^2}.\nonumber
\label{eq:gamma3d}
\end{eqnarray}
\end{widetext}

It is easy to reproduce the ME equations by replacing the full
propagator $G(p)$ appearing in Eq.~(\ref{eq:gamma3qp}) with the free
propagator $G_{0}(p)$, which is equivalent to the bare vertex
approximation $\Gamma_3\rightarrow \sigma_3$. Alternatively, one
could substitute $A_1=A_2=1$ and $\Delta = 0$ into $\Gamma_{30}$,
$\Gamma_{33}$, $\Gamma_{31}$, and $\Gamma_{3d}$, and then obtain
\begin{eqnarray}
\Gamma_{30}=0, \quad \Gamma_{33}=1, \quad \Gamma_{31}=0, \quad
\Gamma_{3d}=0.
\end{eqnarray}
Such manipulations reduce Eqs.~(\ref{eq:A1p}-\ref{eq:Deltap}) to the
standard ME equations (not shown explicitly).

The self-consistent integral equations of
$A_1(\epsilon_{n},\mathbf{p})$, $A_2(\epsilon_{n},\mathbf{p})$, and
$\Delta(\epsilon_{n},\mathbf{p})$ can be numerically solved using
the iteration method (see Ref.~\cite{Liu21} for a detailed
illustration of this method). The computational time required to
reach convergent results depends crucially on the number of integral
variable: adding one variable leads to an exponential increase of
the computational time. The coupled equations
(\ref{eq:A1p}-\ref{eq:Deltap}) have only one variable $\epsilon_{n}$
if all electrons are assumed to reside exactly on the Fermi surface.
Such an assumption simplifies the equations and dramatically
decreases the computational time, but might not be justified in the
present case due to the strong momentum dependence of EPI coupling
strength. Therefore, here we consider all the possible values of
$\mathbf{p}$ and directly solve Eqs.~(\ref{eq:A1p}-\ref{eq:Deltap})
without introducing further approximations. But these equations have
three integral variables, namely $\omega_{m}$, $q_{1}$, and $q_{2}$.
Solving them would consume too many computational resources.

The burden of numerical computation can be greatly lightened if the
number of integral variable is reduced. We suppose that the system
is isotropic and then make an effort to integrate over the angle
$\theta$ between $\mathbf{p}$ and $\mathbf{q}$ before starting the
iterative process. After doing so, only two free variables, namely
$\omega_{m}$ and $|\mathbf{q}|$, are involved in the process of
performing iterations. The computational time can thus be greatly
shortened. Generically, it is not easy to integrate over $\theta$.
In our case, however, although the current vertex function
$\Gamma_{3}(q,p)$ is complicated, the free propagators $D_{0}(q)$
and $F_{0}(q)$ are simple functions of their variables. The phonon
energy $\Omega_{\mathbf{q}}$ is a constant, thus $D_{0}(q)$ depends
solely on the frequency, i.e., $D_{0}(q)=D_{0}(\omega_{m})$. In
comparison, $F_{0}(\mathbf{q})$ depends solely on the momentum. To
illustrate why $\theta$ can be integrated out, it is more convenient
to deal with the DS equation shown in Eq.~(\ref{eq:dsefinal})
instead of the formally complicated equations
(\ref{eq:A1p}-\ref{eq:Deltap}). With the redefinitions
$\mathbf{p+q}\rightarrow \mathbf{k}$ and $\int d\mathbf{q}
\rightarrow \int d\mathbf{k}$, we can re-write
Eq.~(\ref{eq:dsefinal}) as
\begin{eqnarray}
G^{-1}(\epsilon_n,\mathbf{p}) &=& G_{0}^{-1}(\epsilon_n,\mathbf{p})
+ T\sum_m\int \frac{k dk d\theta}{(2\pi)}\sigma_3
G(\epsilon_m,\mathbf{k}) \nonumber \\
&\times& \Big[g^2 D_0(\omega_m) + \frac{e^2}{\varepsilon
\sqrt{\mathbf{p}^{2}+\mathbf{k}^{2} - 2|\mathbf{p}|
|\mathbf{k}|\cos\theta}} \Big] \nonumber \\
&\times& \Gamma_3(\epsilon_n,\mathbf{p},
\epsilon_n+\omega_{m},\mathbf{k}).
\end{eqnarray}
Both $G_{0}(\epsilon_n,\mathbf{p})$ and $G(\epsilon_n,\mathbf{p})$
are independent of $\theta$, thus $\theta$ is not involved in the
iterative process and can be numerically integrated at each step.

$F_{0}(\mathbf{q})$ is singular at $\mathbf{q}=0$, reflecting the
long-range nature of bare Coulomb interaction. If the electrons are
treated by the jellium model, the contribution of $\mathbf{q}=0$
must be eliminated since it cancels out the static potential between
negative and positive charges. This can be implemented by
introducing an infrared cutoff $\delta$. In our calculations, we set
$\delta = 10^{-6}{~p_{\mathrm{F}}^{}}$. We have already confirmed
that the final results of $T_{c}$ are nearly unchanged as $\delta$
varies within the range of $\left[10^{-8}{~p_{\mathrm{F}}^{}},
10^{-3}{~p_{\mathrm{F}}^{}}\right]$. Apart from the infrared cutoff,
it is also necessary to introduce an ultraviolet cutoff $\Lambda$
for the momentum. A natural choice is to set $\Lambda =
p_{\mathrm{F}}^{}$. Our final results are also insensitive to other
choices of $\Lambda$, which might be attributed to the dominance of
small-$\mathbf{q}$ processes.

To facilitate numerical calculations, it is more convenient to make
all the variables to become dimensionless. Dimensional parameters
can be made dimensionless after performing the following re-scaling
transformations:
\begin{eqnarray}
&&\frac{p}{p_{\mathrm{F}}^{}}\rightarrow p, \quad
\frac{k}{p_{\mathrm{F}}^{}}\rightarrow k, \quad
\frac{q}{p_{\mathrm{F}}^{}}\rightarrow q, \quad
\frac{q_{\mathrm{p}}}{p_{\mathrm{F}}^{}}\rightarrow q_{\mathrm{p}}, \\
&& \frac{T}{E_{\mathrm{F}}}\rightarrow T, \quad
\frac{\epsilon_n}{E_{\mathrm{F}}} \rightarrow \epsilon_n, \quad
\frac{\omega_m}{E_{\mathrm{F}}}
\rightarrow\omega_{m}, \\
&& \frac{\Omega_{q}}{E_{\mathrm{F}}} \rightarrow \Omega_{q}, \quad
\frac{\xi_p}{E_{\mathrm{F}}}\rightarrow \xi_{p}, \quad
\frac{g}{E_{\mathrm{F}}} \rightarrow g.
\end{eqnarray}
The parameters $\lambda$ and $\alpha$ are already made dimensionless
and thus kept unchanged. The integral interval of the re-scaled
variable $k$ is $[10^{-6},1]$.

\begin{widetext}

\begin{figure}[H]
\centering
\includegraphics[width=2.3in]{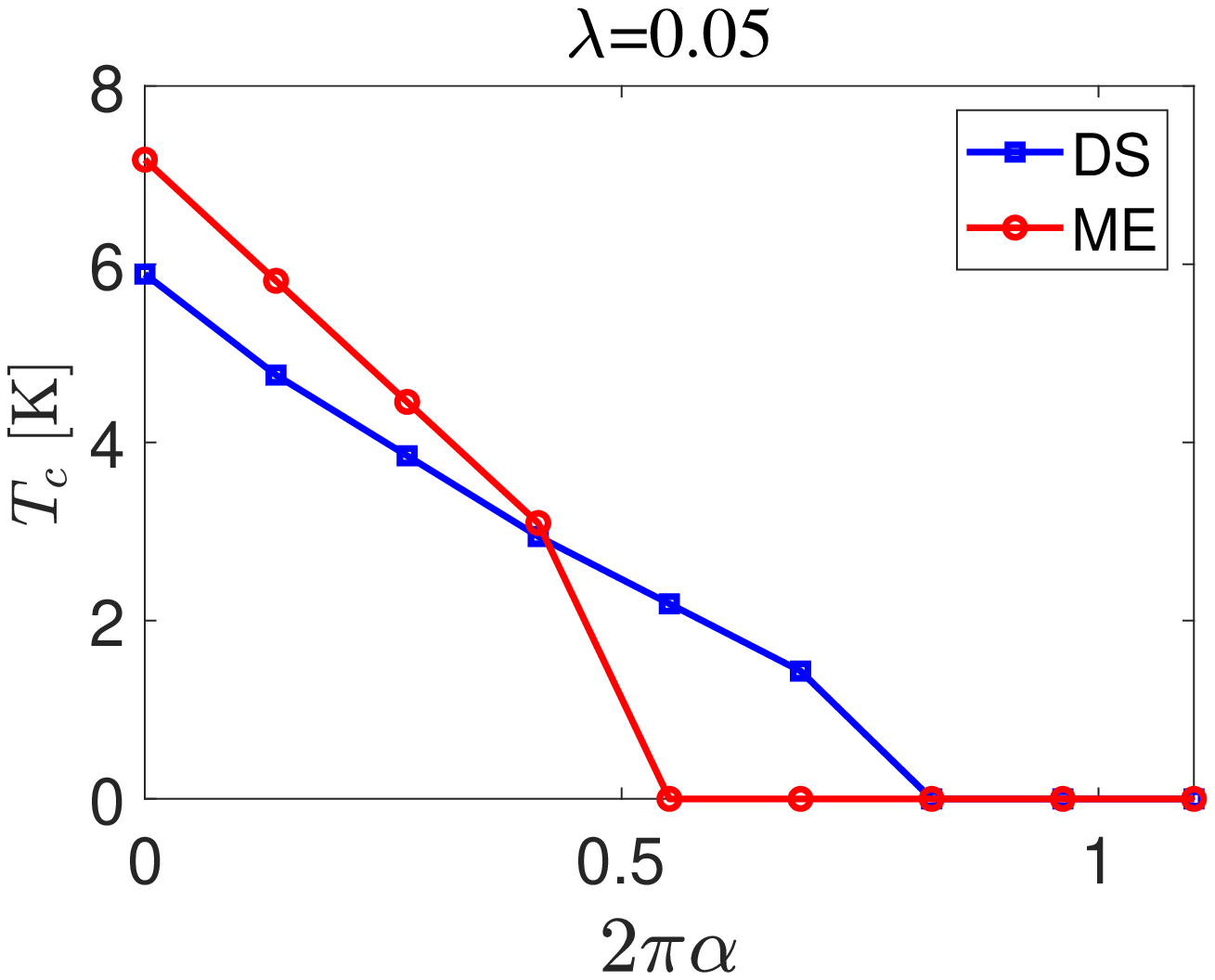}
\includegraphics[width=2.3in]{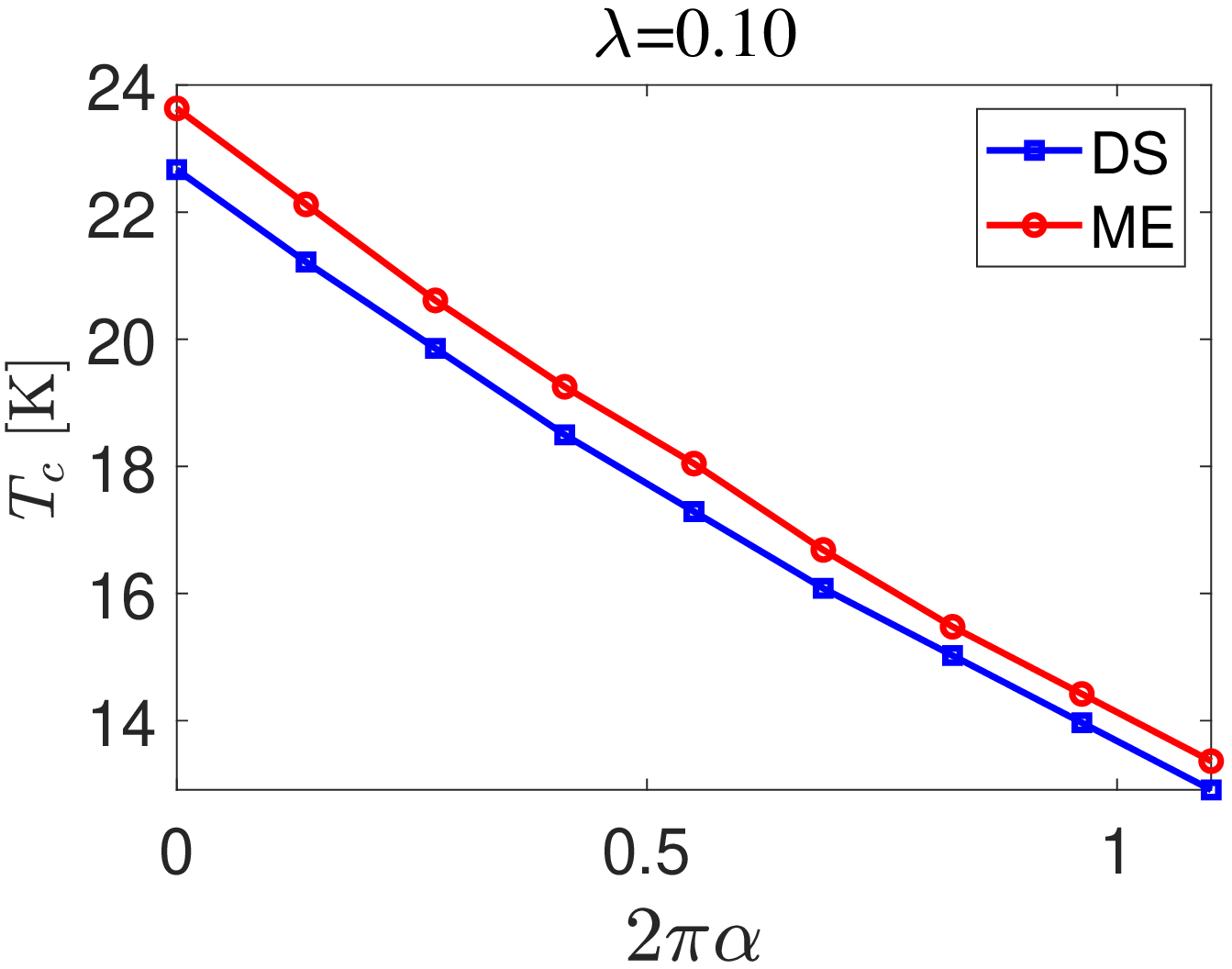}
\includegraphics[width=2.3in]{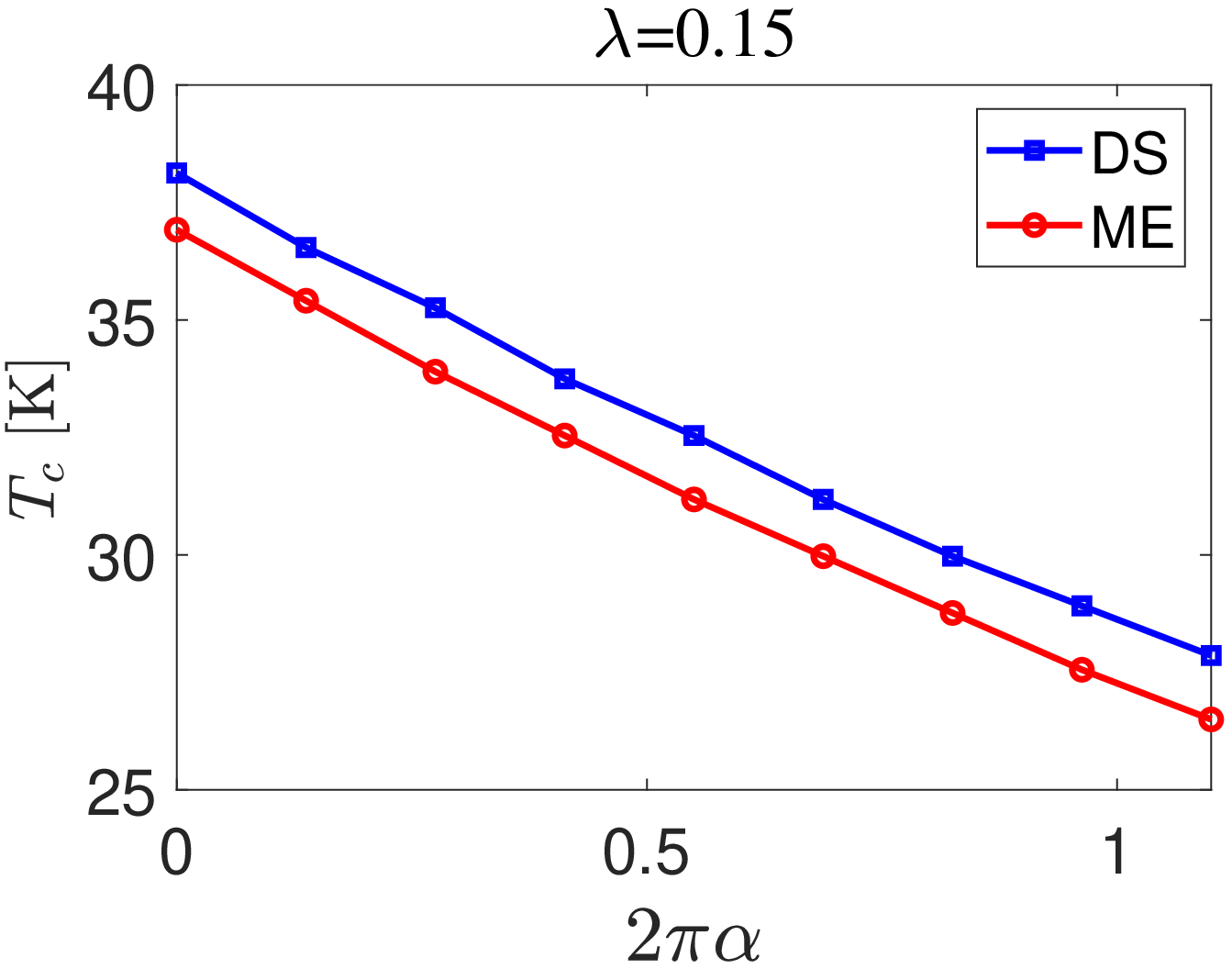}
\includegraphics[width=2.3in]{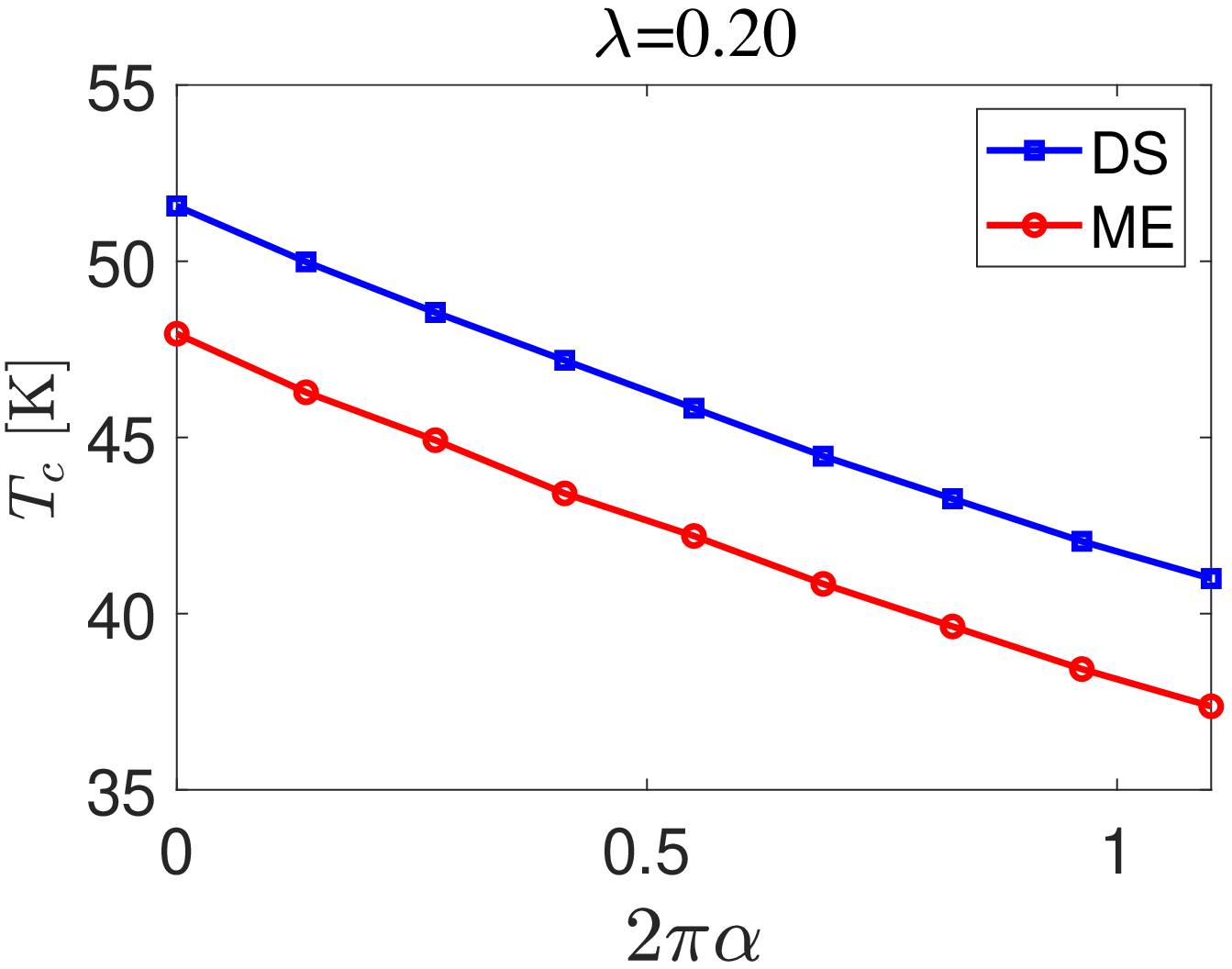}
\includegraphics[width=2.3in]{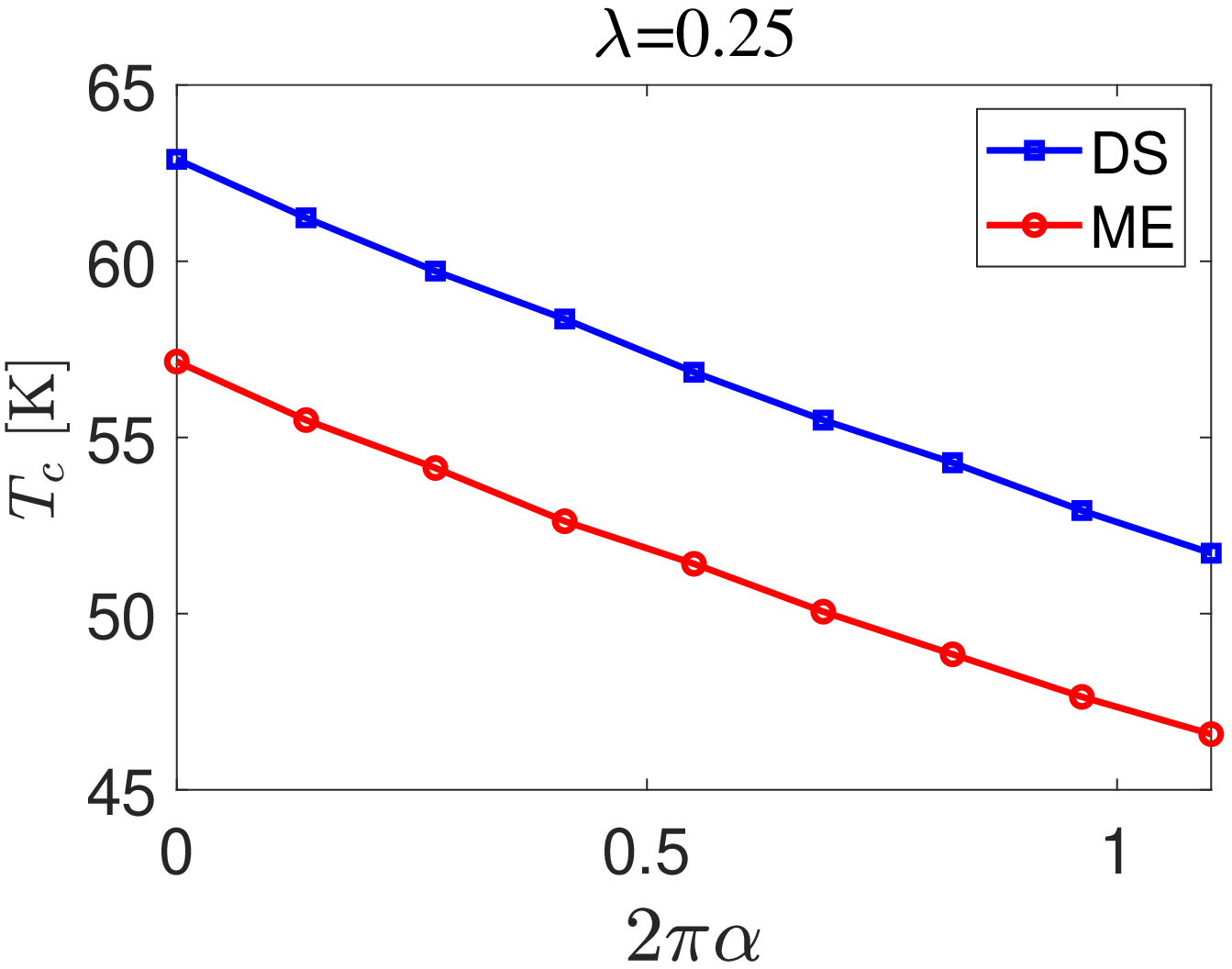}
\includegraphics[width=2.3in]{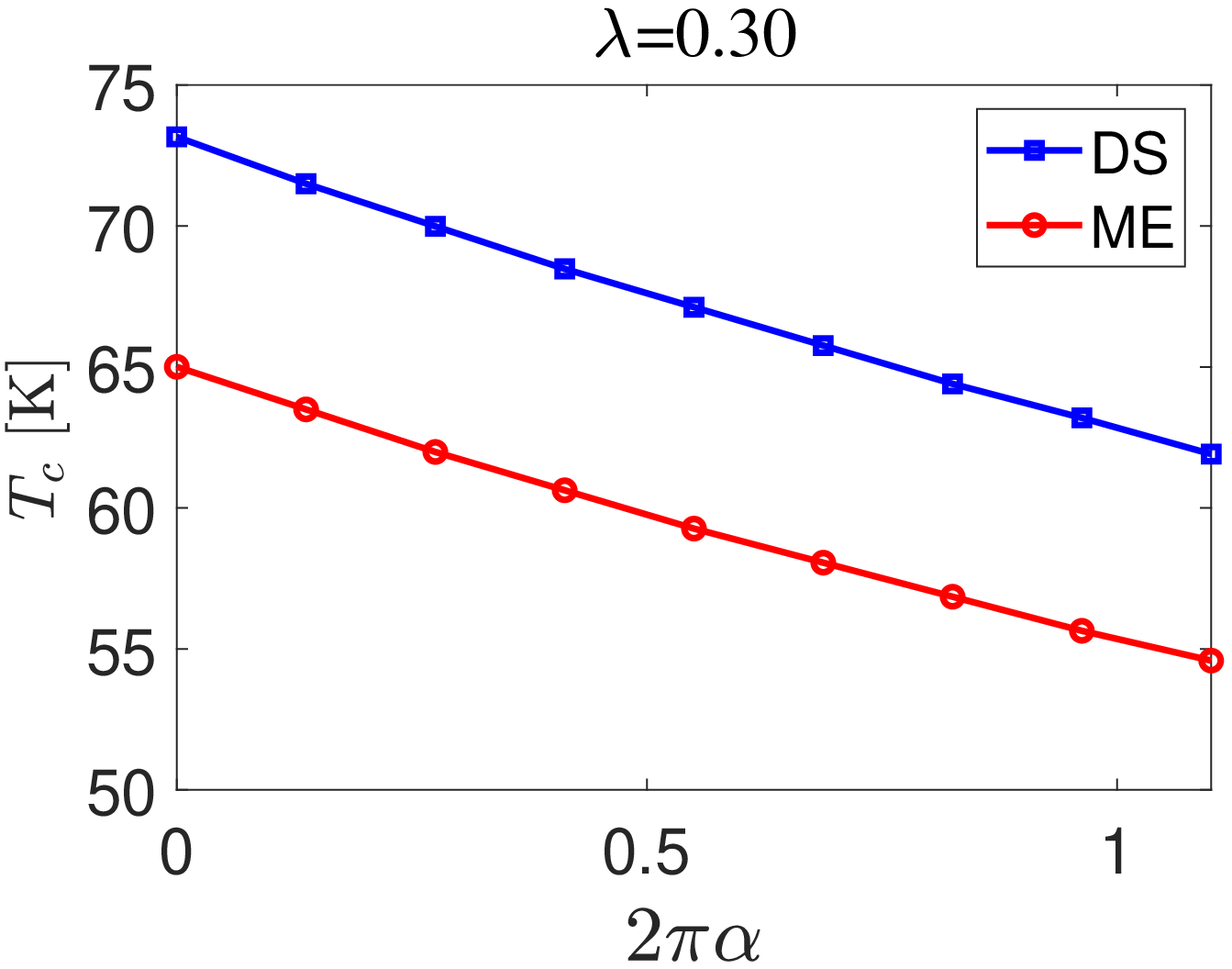}
\caption{Comparison between results of $T_{c}$ obtained by solving
ME and DS equations with six different values of $\lambda$.}
\label{figure}
\end{figure}

\end{widetext}

After making re-scaling transformations, the electron dispersion
$\xi_{\mathbf{p}} = \frac{\mathbf{p}^{2}}{2m_{e}} -
\mu_{\mathrm{F}}^{}$ is turned into $\mathbf{p}^{2}-1$, which is
dimensionless. The resulting integral equations of $A_{1}$, $A_{2}$,
and $\Delta$ do not explicitly depend on neither $p_{\mathrm{F}}^{}$
nor $m_{e}$. Thus, it is not necessary to separately specify the
values of $p_{\mathrm{F}}^{}$ and $m_{e}$, since the final results
of critical temperature only exhibit a dependence on $E_{\mathrm{F}}
= \frac{p_{\mathrm{F}}^{2}}{2m_{e}}$. From the numerical solutions
of DS and ME equations, we could obtain an effective dimensionless
transition temperature, denoted by $T_{c}'$, that is equal to
$T_{c}/E_{\mathrm{F}}$. The Fermi energy $E_{\mathrm{F}} = 65
\mathrm{~meV}$ amounts to approximately $\sim 755\mathrm{~K}$. Then
the actual transition temperature $T_{c}$ can be readily obtained
from $T_{c}'$ through the relation $T_{c} \sim T_{c}'\times
(755\mathrm{~K})$.

It should be emphasized that the free phonon propagator $D_{0}(q)$
is used in both the DS-level and ME-level calculations. Thus we are
allowed to determine the influence of EPI vertex corrections by
comparing the values of $T_{c}$ obtained under these two
approximations. The pairing gap $\Delta$ is supposed to have an
isotropic $s$-wave symmetry \cite{Johnston-NJP2016}. To make our
analysis more generic, we consider six different values of the
strength parameter $\lambda$, including $\lambda=0.05$,
$\lambda=0.10$, $\lambda=0.15$, $\lambda=0.20$, $\lambda=0.25$, and
$\lambda=0.30$. The numerical results of $T_{c}$ are presented in
Fig.~\ref{figure}, where the red and blue curves correspond to the
ME and DS results, respectively.

We first consider the simplest case in which the Coulomb interaction
is absent. In a previous work \cite{Liu21}, it was found that
including EPI vertex corrections tends to promote $T_{c}$ evaluated
at the ME-level (bare vertex). This conclusion was reached based on
the assumption that the electrons always strictly reside on the
Fermi surface such that $\xi_{\mathbf{p}}=\xi_{\mathbf{p}_{F}}=0$
\cite{Liu21}. Here we re-solve Eqs.~(\ref{eq:A1p}-\ref{eq:Deltap})
without making this assumption. From the numerical results presented
in Fig.~\ref{figure}, we observe that the impact of EPI vertex
corrections on $T_{c}$ is strongly dependent of the value of EPI
strength parameter $\lambda$. Specifically, we find that vertex
corrections slightly reduce $T_{c}$ for $\lambda=0.10$, but
considerably enhance $T_{c}$ for $\lambda=0.15$, $\lambda=0.20$,
$\lambda=0.25$, and $\lambda=0.30$. The enhancement of $T_{c}$ due
to vertex corrections becomes more significant as $\lambda$ further
increases. The case of $\lambda=0.05$ appears to be peculiar: the
vertex corrections play different roles as the effective strength of
Coulomb interaction is changed.

%\begin{widetext}

%\begin{figure}[htbp]
%\begin{minipage}[t]{0.3\linewidth}
%\includegraphics[height=4.6cm,width=5.9cm]{tc005.eps}
%\end{minipage}%
%\begin{minipage}[t]{0.3\linewidth}
%\includegraphics[height=4.6cm,width=5.9cm]{tc010.eps}
%\end{minipage}
%\begin{minipage}[t]{0.3\linewidth}
%\includegraphics[height=4.6cm,width=5.9cm]{tc015.eps}
%\end{minipage}%

%\begin{minipage}[t]{0.3\linewidth}
%\centering
%\includegraphics[height=4.6cm,width=5.9cm]{tc020.eps}
%\end{minipage}
%\begin{minipage}[t]{0.3\linewidth}
%\centering
%\includegraphics[height=4.6cm,width=5.9cm]{tc025.eps}
%\end{minipage}%
%\begin{minipage}[t]{0.3\linewidth}
%\centering
%\includegraphics[height=4.6cm,width=5.9cm]{tc030.eps}
%\end{minipage}
%\caption{Comparison between results of $T_{c}$ obtained by solving
%ME and DS equations with six different values of $\lambda$. The
%impact of vertex corrections is substantially enhanced as $\lambda$
%increases.} \label{figure}
%\end{figure}
%\end{widetext}

The effect of the Coulomb interaction on $T_{c}$ can be readily
investigated by varying the tuning parameter $\varepsilon$. As
clearly shown by Fig.~\ref{figure}, $T_{c}$ drops monotonously as
$\varepsilon$ decreases. Such a behavior is certainly in accordance
with expectation, since the Coulomb repulsion weakens the effective
attraction between electrons. In the case of $\lambda=0.05$, $T_{c}$
is slightly reduced by vertex corrections for weak Coulomb repulsion
but is enhanced by vertex corrections when the Coulomb repulsion
becomes strong enough. Superconductivity can be completely
suppressed, with $T_{c}\rightarrow 0$, once the effective strength
of Coulomb repulsion exceeds certain threshold. For larger values of
$\lambda$, the Coulomb repulsion has an analogous impact on $T_{c}$.
However, superconductivity could be entirely suppressed only when
the repulsion becomes unrealistically strong.

For any realistic material, $\varepsilon$ takes a specific value, so
does $T_{c}$. $T_{c}$ is completely determined once all the model
parameters are fixed. In a way, our work provides a first-principle
study of the superconducting transition, although the role of
Anderson mechanism remains to be ascertained.

\section{Summary and Discussion \label{Sec:summary}}

In summary, we have performed a non-perturbative study of the
interplay of EPI and Coulomb repulsion by using the DS equation
approach. We have shown that the DS equation of the full electron
propagator $G(p)$ is self-closed provided that all the higher order
corrections to EPI and Coulomb interaction are incorporated via a
number of exact identities. This self-closed DS equation can be
applied to study the superconducting transition beyond the ME
approximation of EPI and the pseudopotential approximation of
Coulomb repulsion. We have employed this approach to evaluate the
pair-braking temperature $T_{c}$ for the interfacial
superconductivity in 1UC FeSe/SrTiO$_{3}$ system and found that the
value of $T_{c}$ could be significantly miscalculated if the vertex
corrections and the momentum-dependence of relevant quantities are
not taken into account in a reliable way.

The calculations of this work ignored several effects that might
change the value of $T_{c}$ and thus need to be improved in the
future. First of all, the simple one-band model studied by us should
be replaced with a realistic multi-band model that embodies the
actual electronic structure \cite{Opp-PRB2018}. The phonon
self-coupling terms \cite{Johnstoncp} are entirely neglected in our
calculations. Including such self-coupling terms invalidates the two
identities given by Eq.~(\ref{eq:d0gamma3}) and
Eq.~(\ref{eq:f0gamma3}). As a consequence, the DS equation of the
electron propagator $G(p)$ can no longer be made self-closed (see
Ref.~\cite{Liu21} and Ref.~\cite{Pan21} for more details). Moreover,
we did not consider the quantum geometry effects \cite{Yanase21},
which could enhance $T_{c}$ to certain extend. In this sense, our
results of $T_{c}$ cannot be directly compared to the experimental
values. The main achievement of our present work is a methodological
advance in the non-perturbative study of the superconducting
transition driven by the interplay of EPI and Coulomb repulsion.

\section*{ACKNOWLEDGEMENTS}

We thank Jing-Rong Wang and Hao-Fu Zhu for helpful discussions. This
work is supported by the Anhui Natural Science Foundation under
grant 2208085MA11.

\newpage

\begin{widetext}

\textbf{Supplementary Material}

\setcounter{section}{0}

\section{A Brief Introduction}

We shall use the functional integral formalism of quantum field
theory to derive the two Ward-Takahashi identities (WTIs) that we
need to determine the self-closed integral equation of the electron
propagator $G(p)$. The basic procedure of the derivation has been
previously outlined in Ref.~\cite{Liu2021Supp}. To make our present
paper self-contained and easier to understand, here we provide more
calculational details.

The partition function of the model describing EPI and Coulomb
interaction has the form
\begin{eqnarray}
Z[J,K,\eta,\eta^{\dag}] =\int D\phi DA D\psi^{\dag} D\psi e^{i\int
dx \mathcal{L}_{T}[\phi,A,\psi^{\dag},\psi]}, \label{eq:Zoriginal}
\end{eqnarray}
where $\int dx \equiv \int d^{3}x = \int dt d^{2}\mathbf{x}$ and the
total Lagrangian density is $\mathcal{L}_{T} = \mathcal{L}+ J\phi+K
A + \psi^{\dag}\eta + \eta^{\dag}\psi $ with $J$, $K$, $\eta$, and
$\eta^{\dag}$ being external sources. $\phi$ is the phonon field and
$A$ is an auxiliary boson field introduced to describe the Coulomb
interaction. In order to study superconducting pairing, it is
convenient to adopt a two-component Nambu spinor $\psi$ to describe
the electrons. The concrete form of $\mathcal{L}$ will be given
later. It will become clear that the symmetry-induced WTI has the
same form irrespective of whether the Coulomb interaction is
included in the Lagrangian density.

For any given $\mathcal{L}_{T}$, the corresponding $Z$ can be
applied to generate various important correlation functions
\cite{PeskinbookSupp, ItzyksonSupp}. We are mainly interested in connected
correlation functions. For this purpose, we also need to use another
generating functional $W$, defined via $Z$ as follows
\begin{eqnarray}
W[J,K,\eta,\eta^{\dag}]= -i\ln Z[J,K,\eta,\eta^{\dag}].
\end{eqnarray}
All the correlation functions to be studied below are connected. The
external resources are taken to vanish ($J \rightarrow 0$,
$K\rightarrow 0$,$\eta\rightarrow 0$, $\eta^{\dag} \rightarrow 0$ )
after functional derivatives are done.

\section{Review of the work of Engelsberg and Schrieffer \label{Sec:WTIES}}

In a seminal paper, Nambu \cite{Nambu60Supp} made a very concise
discussion of the symmetries associated with charge conservation and
spin conservation in a pure EPI system and also briefly analyzed the
corresponding WTIs. Later, Engelsberg and Schrieffer
\cite{Schrieffer1963Supp} provided a more elaborate field-theoretic
analysis of the pure EPI system and, especially, derived a generic
form of the charge-conservation related WTI in the normal state.
Following the scenario proposed by Nambu \cite{Nambu60Supp}, they argued
\cite{Schrieffer1963Supp} that the integral equation of EPI vertex
function obtained under the ladder approximation is gauge-invariant
if the vertex function is connected to the electron propagator via
the WTI when the electron momentum vanishes. Their results are not
satisfied for three reasons. Firstly, the ladder approximation
employed in their analysis is not justified when the EPI is not
weak. Secondly, their WTI in its original form cannot be used to
solve the integral equation of electron propagator and thus is of
little practical value. Thirdly, their analysis is restricted to the
normal state and should be properly generated to treat Cooper
pairing instability.

Before presenting our own work, it is helpful to first show how to
derive the WTI associated with charge conservation based on the
model investigated by Engelsberg and Schrieffer
\cite{Schrieffer1963Supp}. Their analysis were carried out by studying
the Heisenberg equations of motion for the electron and phonon field
operators. Our derivation will be performed within the framework of
functional integral, following the general strategy demonstrated in
Chapter 9 of Ref.~\cite{PeskinbookSupp}.

Engelsberg and Schrieffer \cite{Schrieffer1963Supp} did not consider the
possibility of Cooper pairing and used ordinary spinor $\Psi$,
rather than Nambu spinor, to describe the electrons. The Lagrangian
density defined in terms of $\Psi$ in real space is given by
\begin{eqnarray}
\mathcal{\mathcal{L}} = \Psi^{\dag}(x)\left(i \partial_{x_0} -
\xi_{\nabla}\right)\Psi(x) +\frac{1}{2}A(x)\mathbb{F}(x)A(x)+ \frac{1}{2}\phi^\dag(x)
\mathbb{D}(x)\phi(x) -A(x)\Psi^{\dag}(x)\Psi(x)- g\phi(x){\Psi}^{\dag}(x)\Psi(x).
\label{eq:lagrangianes}
\end{eqnarray}
Here, the time-space coordinate is $x \equiv (x_{0},x_{1},x_{2})$.
$\partial_{x_0}$ represents the partial derivative
$\frac{\partial}{\partial x_{0}}$ and $\xi_{\nabla} =
\frac{-\mathbf{\nabla}^{2}}{2m_{e}}-\mu_{\mathrm{F}}$ with
$\mathbf{\nabla} = \left(\frac{\partial}{\partial
x_{1}},\frac{\partial}{\partial x_{2}}\right)$. The equations of the
free motion of phonons and auxiliary scalar are $\mathbb{D}(x)\phi(x)=0$
and $\mathbb{F}(x)A(x)=0$ respectively. The EPI coupling
parameter $g$ could be a constant or a function of $x$, which does
not affect the final WTI. Notice that the Coulomb interaction is not
included in this Lagrangian density.

Now make the following change to the spinor field $\Psi$:
\begin{eqnarray}
\Psi \rightarrow {\tilde \Psi}=e^{i\chi(x)}\Psi.
\label{eq:u1symmetryphase}
\end{eqnarray}
Here, $\chi(x)$ is supposed to be an arbitrary infinitesimal
function of $x$. The partition function $Z$ should remain the same
if $\Psi$ is replaced with $\tilde \Psi$, namely
\begin{eqnarray}
Z[J,K,\eta,\eta^{\dag}] = \int D\phi  DA D{\tilde \Psi}^{\dag} D{\tilde \Psi}
e^{i\int dx\mathcal{L}_{T}[\phi,A, {\tilde
\Psi}^{\dag},{\tilde\Psi}]}.\label{eq:Ztransformedes}
\end{eqnarray}
Usually EPI in normal metals does not lead to any quantum anomaly
\cite{PeskinbookSupp}, thus the functional integration measure is
invariant under the above transformation, i.e., $D\Psi^{\dag} D\Psi
= D{\tilde \Psi}^{\dag} D{\tilde \Psi}$. Then we find that
\begin{eqnarray}
\int D\phi  DA D{\tilde \Psi}^{\dag} D{\tilde \Psi} \left[e^{i\int dx
\mathcal{L}_{T}[\phi,A,{\tilde \Psi}^{\dag},{\tilde \Psi}]} - e^{i\int
dx \mathcal{L}_{T}[\phi,A,\Psi^{\dag},\Psi]}\right] = 0.
\end{eqnarray}
Since $\chi(x)$ is infinitesimal, this equation implies that
\begin{eqnarray}
\int d^3x D\phi DA\int D{\tilde \Psi}^{\dag} D{\tilde \Psi} ~
\left[\frac{\delta}{\delta \Psi(x)}e^{i\int dx
\mathcal{L}_{T}}(i\chi(x) \Psi(x)) + (i\chi(x)\Psi^{\dag}(x))
\frac{\delta}{\delta \Psi^{\dag}(x)}e^{i\int dx
\mathcal{L}_{T}}\right]=0. \label{eq:u1symmetryes}
\end{eqnarray}
It is easy to verify that
\begin{eqnarray}
\frac{\delta}{\delta \Psi(x)}e^{i\int dx\mathcal{L}_{T}} &=&
ie^{i\int dx \mathcal{L}_{T}}\left[(i \partial_{x_0} + \xi_{\nabla})
\Psi^{\dag}(x) +A(x)\Psi^{\dag}(x)+ g \phi(x)\Psi^{\dag}(x)-
\eta^{\dag}(x)\right], \label{eq:derivepsies}\\
\frac{\delta}{\delta \Psi^{\dag}(x)}e^{i\int dx\mathcal{L}_{T}} &=&
ie^{i\int dx\mathcal{L}_{T}}\left[\left(i \partial_{x_0} -
\xi_{\nabla}\right) \Psi(x) -A(x)\Psi(x)- g \phi(x)\Psi(x) + \eta(x)\right].
\label{eq:derivepsidaggeres}
\end{eqnarray}
Substituting Eq.~(\ref{eq:derivepsies}) and
Eq.~(\ref{eq:derivepsidaggeres}) into Eq.~(\ref{eq:u1symmetryes})
leads to
\begin{eqnarray}
0 &=& \Big\langle \int d^3x ~\chi(x)\Big[\left((i\partial_{x_{0}} +
\xi_{-\nabla})\Psi^{\dag}(x)\right)\Psi(x) + g\phi(x)
\Psi^{\dag}(x)\Psi(x)+A(x)\Psi^{\dag}(x)\Psi(x) -
\eta^{\dag}(x)\Psi(x)\nonumber \\
&& +\Psi^{\dag}(x)(i\partial_{x_0}-\xi_{\nabla})\Psi(x) -
g\phi(x)\Psi^{\dag}(x)\Psi(x)-A(x)\Psi^{\dag}(x)\Psi(x)+
\Psi^{\dag}(x)\eta(x)\Big]\Big\rangle \nonumber\\
&=&\Big\langle\int d^3x~\chi(x)\Big[i\partial_{x_0}
\left(\Psi^{\dag}(x)\Psi(x)\right)+\left(\xi_{-\nabla}
\Psi^{\dag}(x)\right)\Psi(x)-\Psi^{\dag}(x)\left(\xi_{\nabla}
\Psi(x)\right)+\Psi^{\dag}(x)\eta(x)-\eta^{\dag}(x)\Psi(x)
\Big]\Big\rangle .\label{eq:generalcurrent}
\end{eqnarray}
From the above derivations, one could find that
Eq.~(\ref{eq:generalcurrent}) is always correct no matter whether
the model contains only EPI, only Coulomb interaction, or both of
them. In addition, notice that Eq.~(\ref{eq:generalcurrent}) holds
for any continuously differentiable function $\xi_{\nabla}$. In the
simplest case, the dispersion is $\xi_{\nabla} =
\frac{-\mathbf{\nabla}^{2}}{2m_{e}}-\mu_{\mathrm{F}}$. If we
consider a tight-binding model, then the dispersion may be of the
form \cite{Johnston-NJP2016Supp}
$\xi_{\nabla}\equiv\xi({-i\partial_1,-i\partial_2}) =
-t[\cos(-ia\partial_1)+\cos(-ia\partial_2)]-\mu_{\mathrm{F}}$, where
$a$ is lattice constant. In both cases, $\xi_{-\nabla} =
\xi_{\nabla}$. Since $\chi(x)$ is an arbitrary function, the
following equation should be obeyed
\begin{eqnarray}
\big\langle\big[i\partial_{x_0}\left(\Psi^{\dag}(x)
\Psi(x)\right)+\left(\xi_{\nabla}\Psi^{\dag}(x)\right) \Psi(x)
-\Psi^{\dag}(x)\left(\xi_{\nabla}\Psi(x)\right)\big]\big\rangle &=&
\langle \eta^{\dag}(x)\Psi(x)\rangle -
\langle\Psi^{\dag}(x)\eta(x)\rangle.
\label{eq:chargeSTIordinaryspinor}
\end{eqnarray}

To simplify notations and also to make a direct comparison to the
work of Ref.~\cite{Schrieffer1963Supp}, in the following we shall take
the simplest dispersion $\xi_{\nabla} =
\frac{-\mathbf{\nabla}^{2}}{2m_{e}}-\mu_{\mathrm{F}}$ as an example
to explain how the WTI is obtained. The generic expression of WTI
has the same form if other choices of $\xi_{\nabla}$ are made. The
left-hand side of Eq.~(\ref{eq:chargeSTIordinaryspinor}) can be
identified as the vacuum expectation value of the divergence of the
composite current operator $j_{\mu}(x)\equiv
\left(j_{0}(x),\mathbf{j}(x)\right)$, where
\begin{eqnarray}
j_{0}(x) &=& \Psi^{\dag}(x)\Psi(x), \\
\mathbf{j}(x) &=& \frac{1}{2m_{e}}\big[\left(i\mathbf{\nabla}
\Psi^{\dag}(x)\right)\Psi(x) - \Psi^{\dag}(x)
\left(i\mathbf{\nabla}\Psi(x)\right)\big].
\label{eq:schriefferccurrent}
\end{eqnarray}
This identification can be readily confirmed since
\begin{eqnarray}
\langle i\partial_{\mu}j_{\mu}(x)\rangle &=& \langle
i \partial_{x_0}j_{0}(x)\rangle + \langle
i\mathbf{\nabla}\cdot \mathbf{j}(x)\rangle \nonumber \\
&=& \big\langle i \partial_{x_0}\left(\Psi^{\dag}(x)\Psi(x)\right)
\big\rangle + \frac{1}{2m_{e}}\big\langle i\mathbf{\nabla}\cdot
\left[\left(i\mathbf{\nabla} \Psi^{\dag}(x)\right)\Psi(x) -
\Psi^{\dag}(x)\left(i\mathbf{\nabla} \Psi(x)\right)\right]
\big\rangle \nonumber \\
&=& \big\langle i \partial_{x_0}\left(\Psi^{\dag}(x)\Psi(x)\right)
\big\rangle + \big\langle\left(\xi_{\nabla}
\Psi^{\dag}(x)\right)\Psi(x) - \Psi^{\dag}(x)
\left(\xi_{\nabla}\Psi(x)\right)\big\rangle.
\end{eqnarray}
Now the identity given by Eq.~(\ref{eq:chargeSTIordinaryspinor}) can
be re-written as
\begin{eqnarray}
\langle i \partial_{x_0}j_{0}(x) + i\mathbf{\nabla}\cdot
\mathbf{j}(x)\rangle =  \langle \eta^{\dag}(x)\Psi(x)\rangle -
\langle \Psi^{\dag}(x)\eta(x)\rangle.\label{eq:STIchargees}
\end{eqnarray}
If all the external resources are eliminated by taking the limit of
$\eta^{\dag}=\eta=0$, this identity re-produces the famous Noether
theorem, i.e.,
\begin{eqnarray}
\langle i \partial_{x_0}j_{0}(x) + i\mathbf{\nabla}\cdot
\mathbf{j}(x)\rangle = 0.
\end{eqnarray}
Nevertheless, it is more advantageous to keep all the external
sources at the present stage. Actually, one could obtain very
important results if one performs functional derivatives
$\frac{\delta}{\delta\eta(z)}$ and
$\frac{\delta}{\delta\eta^{\dag}(y)}$ to both sides of
Eq.~(\ref{eq:STIchargees}) in order before taking external sources
to zero. This manipulation drives the right-hand side of
Eq.~(\ref{eq:STIchargees}) to become
\begin{eqnarray}
\frac{\delta}{\delta \eta^{\dag}(y)}\frac{\delta}{\delta \eta(z)}
\left[\langle \eta^{\dag}(x)\Psi(x)\rangle - \langle
\Psi^{\dag}(x)\eta(x)\rangle\right] &=& \frac{\delta}{\delta
\eta^{\dag}(y)}\frac{\delta}{\delta \eta(z)}
\left[\eta^{\dagger}(x)\frac{\delta W}{\delta
\eta^{\dagger}(x)}+\frac{\delta W}{\delta\eta(x)}\eta(x)\right]
\nonumber \\
&=&-\delta(y-x)\frac{\delta W}{\delta\eta(z)\delta
\eta^{\dagger}(x)}-\frac{\delta W}{\delta \eta^{\dag}(y)
\delta\eta(x)}\delta(x-z),
\end{eqnarray}
which can be readily identified as
\begin{eqnarray}
\delta(y-x)G(x-z)-G(y-x)\delta(x-z).\label{eq:rhsofwti}
\end{eqnarray}
In the above derivation, we have used the following identities:
\begin{eqnarray}
&&\frac{\delta W}{\delta\eta(z)}\Big|_{\eta,\eta^{\dag},J=0} =
-\langle\Psi^{\dagger}(z)\rangle, \quad \frac{\delta
W}{\delta\eta^{\dagger}(y)}\Big|_{\eta,\eta^{\dag},J=0} =
\langle\Psi(y)\rangle, \nonumber \\
&&\frac{\delta^2 W}{\delta \eta^{\dagger}(y) \delta\eta(z)}
\Big|_{\eta,\eta^{\dag},J=0}=G(y-z).
\end{eqnarray}
The same manipulation turns the right-hand side of
Eq.~(\ref{eq:STIchargees}) into
\begin{eqnarray}
&& \frac{\delta}{\delta \eta^{\dag}(y)}\frac{\delta}{\delta \eta(z)}
\langle i \partial_{x_0} j_{0}(x) + i\mathbf{\nabla} \cdot
\mathbf{j}(x)\rangle \nonumber \\
&=& \langle i \partial_{x_0} j_{0}(x)\Psi(y)\Psi^{\dag}(z)\rangle +
\langle i \mathbf{\nabla}\cdot
\mathbf{j}(x)\Psi(y)\Psi^{\dag}(z)\rangle \nonumber \\
&=& i \partial_{x_0}\big\langle \Psi^{\dag}(x)\Psi(x)\Psi(y)
\Psi^{\dag}(z)\big\rangle + \frac{1}{2m_{e}}i\mathbf{\nabla}\cdot
\big\langle\big[\left(i\mathbf{\nabla} \Psi^{\dag}(x)\right)\Psi(x)
- \Psi^{\dag}(x)\left(i\mathbf{\nabla} \Psi(x)\right)\big]\Psi(y)
\Psi^{\dag}(z)\big\rangle. \label{eq:partialjmupsipsidag}
\end{eqnarray}
Here, notice an important feature that the derivatives
$\partial_{x_0}$ and $\mathbf{\nabla}$ can be freely moved out of
the angle-bracket $\langle \cdots \rangle$ when the mean value is
defined within the framework of function integral \cite{PeskinbookSupp}.

Now define a scalar function ${\tilde \Gamma}_{0}$ and a vector
function ${\tilde {\mathbf{\Gamma}}} = ({\tilde\Gamma}_{1},{\tilde
\Gamma}_{2})$ as follows
\begin{eqnarray}
\big\langle \Psi^{\dag}(x)\Psi(x)\Psi(y)\Psi^{\dag}(z)\big\rangle =
-\int d\zeta d\zeta' G(y-\zeta){\tilde \Gamma}_{0}(\zeta-x,x-\zeta')
G(\zeta'-z),\label{eq:currentvertexgamma0} \\
\frac{1}{2m_{e}}\big\langle\big[\left(i\mathbf{\nabla}
\Psi^{\dag}(x)\right) \Psi(x) -
\Psi^{\dag}(x)\left(i\mathbf{\nabla}\Psi(x)\right)\big]
\Psi(y)\Psi^{\dag}(z)\big\rangle = -\int d\zeta d\zeta'
G(y-\zeta){\tilde {\mathbf{\Gamma}}}(\zeta-x,x-\zeta') G(\zeta'-z),
\label{eq:currentvertexgammavector}
\end{eqnarray}
Functions ${\tilde \Gamma}_{0}$ and ${\tilde {\mathbf{\Gamma}}}$ are
called current vertex functions \cite{Liu2021Supp} because they are
defined in terms of the time-component $j_{0}(x)$ and the spatial
component $\mathbf{j}(x)$ of the composite current operator
$j_{\mu}(x)$, respectively. Then Eq.~(\ref{eq:partialjmupsipsidag})
is expressed in terms of ${\tilde \Gamma}_{0}$ and
${\tilde{\mathbf{\Gamma}}}$ as
\begin{eqnarray}
-\int d\zeta d\zeta' G(y-\zeta)i \partial_{x_0}{\tilde
\Gamma}_{0}(\zeta-x, x-\zeta')G(\zeta'-z) -\int d\zeta d\zeta'
G(y-\zeta)i\mathbf{\nabla}\cdot
{\tilde{\mathbf{\Gamma}}}(\zeta-x,x-\zeta')G(\zeta'-z),
\label{eq:lhsofwti}
\end{eqnarray}
where
\begin{eqnarray}
i \partial_{x_0}{\tilde \Gamma}_{0}(\zeta-x, x-\zeta') &\equiv&
i\frac{\partial}{\partial x_{0}}{\tilde \Gamma}_{0}(\zeta-x,
x-\zeta'), \\
i\mathbf{\nabla}\cdot {\tilde \Gamma}_{0}(\zeta-x, x-\zeta')
&\equiv& i\frac{\partial}{\partial x_{1}}{\tilde
\Gamma}_{1}(\zeta-x, x-\zeta') + i\frac{\partial}{\partial
x_{2}}{\tilde \Gamma}_{2}(\zeta-x, x-\zeta').
\end{eqnarray}

The two expressions given by Eq.~(\ref{eq:rhsofwti}) and
Eq.~(\ref{eq:lhsofwti}) are equal since they both stem from
Eq.~(\ref{eq:STIchargees}), namely
\begin{eqnarray}
&& -\int d\zeta d\zeta' G(y-\zeta)\big[i\partial_{x_0}{\tilde
\Gamma}_{0}(\zeta-x,x-\zeta')+i\mathbf{\nabla}\cdot
{\tilde{\mathbf{\Gamma}}}(\zeta-x,x-\zeta')\big]G(\zeta'-z) \nonumber \\
&=& \delta(y-x)G(x-z)-G(y-x)\delta(x-z). \label{eq:wtiesrealspace}
\end{eqnarray}
This is actually the real-space WTI. Its expression can be made less
awkward by performing Fourier transformations. The Fourier
transformation of the right-hand side of
Eq.~(\ref{eq:wtiesrealspace}) is very simple and the result is
\begin{eqnarray}
\delta(y-x)G(x-z)-G(y-x)\delta(x-z) = \int \frac{d^3p}{(2\pi)^3}
\frac{d^3q}{(2\pi)^3} \left[G(p)-G(p+q)\right]
e^{-i(p+q)(y-x)-ip(x-z)}.\label{eq:Fouriertransformgpes}
\end{eqnarray}
The Fourier transformation of the left-hand side is a little more
complicated, and can be carried out step by step. Let us take $
\partial_{x_0}\langle\Psi^{\dag}(x)\Psi(x)\Psi(y)\Psi^{\dag}(z)
\rangle$ as an example and show how the transformation is
implemented below:
\begin{eqnarray}
&&i \partial_{x_0} \langle \Psi^{\dag}(x)\Psi(x)\Psi(y)
\Psi^{\dag}(z)\rangle \nonumber \\
&=& -\int d\zeta d\zeta' G(y-\zeta)i \partial_{x_0}{\tilde
\Gamma}_{0}(\zeta-x,x-\zeta')G(\zeta'-z) \nonumber \\
&=& -i \partial_{x_0}\int d\zeta d\zeta'
\int\frac{d^3p_{1}}{(2\pi)^3} G(p_{1})e^{-i p_{1}(y-\zeta)}\int
\frac{d^3q}{(2\pi)^3}
\frac{d^3p}{(2\pi)^3}e^{-i(p+q)(\zeta-x)}{\tilde
\Gamma}_{0}(q,p)e^{-ip(x-\zeta')} \int
\frac{d^3p_{2}}{(2\pi)^3}G(p_{2})e^{-ip_{2}(\zeta'-z)}
\nonumber \\
&=& -i \partial_{x_0}\int \frac{d^3p}{(2\pi)^3}\frac{d^3q}{(2\pi)^3}
\left[\int \frac{d^3p_{1}}{(2\pi)^3}\frac{d^3p_{2}}{(2\pi)^3} d\zeta
d\zeta' e^{i\left(p_{1}-(p+q)\right)\zeta}
e^{i(p-p_{2})\zeta'}\right] G(p_{1}){\tilde
\Gamma}_{0}(q,p)G(p_{2})e^{iqx-ip_{1}y+ip_{2}z}
\nonumber\\
&=& -i \partial_{x_0}\int \frac{d^3p}{(2\pi)^3}
\frac{d^3q}{(2\pi)^3}\left[\int \frac{d^3p_{1}}{(2\pi)^3}
\frac{d^3p_{2}}{(2\pi)^3}(2\pi)^6
\delta(p_{1}-(p+q))\delta(p-p_{2})\right]G(p_{1}){\tilde
\Gamma}_{0}(q,p)G(p_{2}) e^{iqx - ip_{1}y + ip_{2}z}
\nonumber \\
&=& \int\frac{d^3p}{(2\pi)^3}\frac{d^3q}{(2\pi)^3}G(p+q)q_{0}
{\tilde \Gamma}_{0}(q,p)G(p)e^{-i(p+q)(y-x)}e^{-ip(x-z)}.
\label{eq:Fouriertransform1es}
\end{eqnarray}
Here, the electron propagator $G(y-\zeta)$ and the functions
${\tilde \Gamma}_{0}(\zeta-x,x-\zeta')$ and
${\tilde{\mathbf{\Gamma}}}(\zeta-x,x-\zeta')$ have been Fourier
transformed according to the expressions presented in Appendix B of
Ref.~\cite{Schrieffer1963Supp}. Following the same procedure, it is
straightforward to obtain
\begin{eqnarray}
\langle i\mathbf{\nabla} \cdot \mathbf{j}(x)\Psi(y)
\Psi^{\dag}(z)\rangle &=& \frac{1}{2m_{e}} i\mathbf{\nabla}\cdot
\big\langle\big[\left(i\mathbf{\nabla}\Psi^{\dag}(x)\right)\Psi(x) -
\Psi^{\dag}(x)\left(i\mathbf{\nabla} \Psi(x)\right)\big]\Psi(y)
\Psi^{\dag}(z)\big\rangle \nonumber \\
&=& -\int d\zeta d\zeta' G(y-\zeta)i\mathbf{\nabla}\cdot
{\tilde{\mathbf{\Gamma}}}(\zeta-x,x-\zeta')G(\zeta'-z) \nonumber \\
&=& -\int \frac{d^3p}{(2\pi)^3}\frac{d^3q}{(2\pi)^3}
G(p+q)\mathbf{q}\cdot {\tilde {\mathbf \Gamma}}(q,p)G(p)
e^{-i(p+q)(y-x)-ip(x-z)}. \label{eq:Fouriertransform2es}
\end{eqnarray}

Combining the results given by Eq.~(\ref{eq:Fouriertransform1es}),
Eq.~(\ref{eq:Fouriertransform2es}), and
Eq.~(\ref{eq:Fouriertransformgpes}) leads to
\begin{eqnarray}
G(p+q)\left[q_{0}{\tilde \Gamma}_{0}(q,p) - \mathbf{q}\cdot {\tilde
{\mathbf \Gamma}}(q,p)\right]G(p) = G(p) - G(p+q).
\end{eqnarray}
This identity can be readily changed into a more compact form
\begin{eqnarray}
q_{0}{\tilde \Gamma}_{0}(q,p) - \mathbf{q}\cdot {\tilde {\mathbf
\Gamma}}(q,p) = G^{-1}(p+q)-G^{-1}(p). \label{eq:chargewties}
\end{eqnarray}
This is precisely the WTI obtained previously by Engelsberg and
Schrieffer \cite{Schrieffer1963Supp}. In order to determine the DS
equation of electron propagator $G(p)$, it is necessary to first get
the scalar function ${\tilde \Gamma}_{0}(q,p)$. However, it is
apparently not possible to determine ${\tilde \Gamma}_{0}(q,p)$ by
solving one single WTI, since the vector function ${\tilde {\mathbf
\Gamma}}(q,p)$ is not known. Engelsberg and Schrieffer
\cite{Schrieffer1963Supp} did not try to made any effort to explore the
structure of ${\tilde {\mathbf \Gamma}}(q,p)$. They simply assumed
that $\mathbf{q}\cdot {\tilde {\mathbf \Gamma}}(q,p) = 0$ as the
limit $\mathbf{q} \rightarrow 0$ is taken and then simplified the
WTI given by Eq.~(\ref{eq:chargewties}) into $q_{0}{\tilde
\Gamma}_{0}(q,p) = G^{-1}(p+q)-G^{-1}(p)$ in this limit.

\section{Two coupled Ward-Takahashi identities in Nambu spinor representation \label{Sec:WTIsnambuspinor}}

In order to investigate Cooper pairing, here we adopt two-component
Nambu spinor $\psi$ to describe electrons. The Lagrangian density
defined in terms of $\psi$ is given by
\begin{eqnarray}
\mathcal{\mathcal{L}} &=& \psi^{\dag}(x)\left(i\partial_{x_0}
\sigma_{0} - \xi_{\nabla}\sigma_{3}\right)\psi(x) +
\frac{1}{2}A(x)\mathbb{F}(x)A(x)+\frac{1}{2}\phi^\dag(x)
\mathbb{D}(x)\phi(x) -A(x)\psi^{\dag}(x)\sigma_{3}\psi(x)\nonumber\\
&-& g\phi(x){\psi}^{\dag}(x)\sigma_{3}\psi(x).
\label{eq:lagrangiannambuspinor}
\end{eqnarray}
The partition function $Z$ for this model is invariant under the
following two different transformations:
\begin{eqnarray}
&&\psi \rightarrow {\tilde \psi}=e^{i\chi(x) \sigma_{3}}\psi,
\label{eq:u1charge} \\
&& \psi \rightarrow {\tilde \psi}=e^{i\chi(x) \sigma_{0}}\psi.
\label{eq:u1spin}
\end{eqnarray}
These two transformations can be uniformly described as
\begin{eqnarray}
\psi \rightarrow {\tilde \psi}=e^{i\chi(x) \sigma_{i}}\psi,
\label{eq:u1symmetrynambu}
\end{eqnarray}
where $i=3,0$. The invariance of $Z$ leads to

\begin{eqnarray}
\int D\phi DA D\psi^{\dag} D\psi \left[e^{i\int dx \mathcal{L}_{T}
[\phi,A,{\tilde \psi}^{\dag},{\tilde \psi}]} - e^{i\int dx
\mathcal{L}_{T}[\phi,A,\psi^{\dag},\psi]}\right] = 0,
\end{eqnarray}
which further leads to
\begin{eqnarray}
\int d^3x\int D\phi DA D\psi^{\dag} D\psi ~ \left[\frac{\delta}{\delta
\psi(x)}e^{i\int dx \mathcal{L}_{T}}(i\sigma_{i}\chi(x) \psi(x)) +
(i\sigma_{i}\chi(x)\psi^{\dag}(x)) \frac{\delta}{\delta
\psi^{\dag}(x)}e^{i\int dx \mathcal{L}_{T}}\right] = 0.
\label{eq:zzequal}
\end{eqnarray}
In the present case, one finds that
\begin{eqnarray}
\frac{\delta}{\delta \psi(x)}e^{i\int dx \mathcal{L}_{T}} &=&
ie^{i\int dx \mathcal{L}_{T}}\left[(i\partial_{x_0}
\psi^{\dag}(x)\sigma_{0} + \xi_{\nabla}\psi^{\dag}(x)\sigma_{3}) +
A(x)\psi^{\dag}\sigma_{3} + g \phi(x)\psi^{\dag}(x)
\sigma_{3}-\eta^{\dag}(x)\right],
\label{eq:derivepsi}\\
\frac{\delta}{\delta \psi^{\dag}(x)}e^{i\int dx \mathcal{L}_{T}} &=&
ie^{i\int dx \mathcal{L}_{T}}\left[\left(i\partial_{x_0}\sigma_{0} -
\xi_{\nabla}\sigma_{3}\right)\psi(x)-A(x)\sigma_{3}\psi(x) -
g\phi(x)\sigma_{3}\psi(x) +\eta(x)\right].
\label{eq:derivepsidagger}
\end{eqnarray}
Inserting Eq.~(\ref{eq:derivepsi}) and
Eq.~(\ref{eq:derivepsidagger}) into Eq.~(\ref{eq:zzequal}) gives
rise to
\begin{eqnarray}
&& \int d^3x\int  D\psi^{\dag} D\psi D\phi ~ e^{i\int dx
\mathcal{L}_{T}} \chi(x)\big[-i \partial_{x_0}(\psi^{\dag}(x)
\sigma_{i}\psi(x)) + \psi^{\dag}(x)(\sigma_{i}\sigma_{3}
\xi_{\nabla})\psi(x) - \xi_{\nabla}\psi^{\dag}(x)
\sigma_{3}\sigma_{i}\psi(x)
\nonumber \\
&& + g\phi(x)\psi^{\dag}(x) [\sigma_{i},\sigma_{3}]\psi(x) +
\eta^{\dag}(x)\sigma_{i}\psi(x) - \psi^{\dag}(x)\sigma_{i}
\eta(x)\big]=0,\label{eq:eq41}
\end{eqnarray}
which, given that $\chi(x)$ is arbitrary, indicates the validity of
the equation
\begin{eqnarray}
\big\langle i\partial_{x_0}(\psi^{\dag}(x) \sigma_{i}\psi(x))
\big\rangle + \big\langle \xi_{\nabla}\psi^{\dag}(x)\sigma_{3}
\sigma_{i}\psi(x)\big\rangle - \big\langle \psi^{\dag}(x)\sigma_{i}
\sigma_{3}\xi_{\nabla}\psi(x)\big \rangle = \langle
\eta^{\dag}(x)\sigma_{i}\psi(x)\rangle - \langle
\psi^{\dag}(x)\sigma_{i}\eta(x)\rangle.\label{eq:STInambu}
\end{eqnarray}
In the above derivation, we have used the relations
$[\sigma_{0},\sigma_{3}] = [\sigma_{3},\sigma_{3}]=0$. Once again,
this equation holds in the presence or absence of the Coulomb
interaction.

For $\sigma_{i}=\sigma_{3}$, the transformation
Eq.~(\ref{eq:u1charge}) corresponds to a conserved current
$j^{c}_{\mu}(x) \equiv (j^{c}_{t}(x),\mathbf{j}^{c}(x))$, where
\begin{eqnarray}
j^{c}_{t}(x) &=& \psi^{\dag}(x)\sigma_{3}\psi(x), \\
\mathbf{j}^{c}(x) &=& \frac{1}{2m_{e}}\big[\left(i\mathbf{\nabla}
\psi^{\dag}(x)\right)\sigma_{0}\psi(x) - \psi^{\dag}(x)\sigma_{0}
\left(i\mathbf{\nabla}\psi(x)\right)\big]. \label{eq:ccurrent}
\end{eqnarray}
Here, notice that $j^{c}_{t}(x)$ and $\mathbf{j}^{c}(x)$ are defined
by two different $2\times 2$ matrices when the two-component Nambu
spinor is used to define the Lagrangian density. This current is
conserved, namely $i\partial_{\mu}j_{\mu}^{c}=0$, in the absence of
external sources, corresponding to the conservation of electric
charge. The generic identity Eq.~(\ref{eq:STInambu}) becomes
\begin{eqnarray}
\big\langle i \partial_{x_0}(\psi^{\dag}(x)
\sigma_{3}\psi(x))\big\rangle + \big\langle
\xi_{\nabla}\psi^{\dag}(x)\sigma_{0}\psi(x) -
\psi^{\dag}(x)\sigma_{0}\xi_{\nabla}\psi(x)\big\rangle = \langle
\eta^{\dag}(x)\sigma_{3}\psi(x)\rangle - \langle
\psi^{\dag}(x)\sigma_{3}\eta(x)\rangle.\label{eq:chargeSTI}
\end{eqnarray}
It is easy to verify that the divergence of conserved current
$j^{c}_{\mu}(x)$ has the form
\begin{eqnarray}
\big\langle i\partial_{\mu}j^{c}_{\mu}(x)\big\rangle &=& \langle i
\partial_{x_0}j^{c}_{t}(x)\rangle + \langle
i\mathbf{\nabla}\cdot \mathbf{j}^{c}(x)\rangle \nonumber \\
&=& \big\langle i\partial_{x_0} \left(\psi^{\dag}(x) \sigma_{3}
\psi(x)\right)\big\rangle + \frac{1}{2m_{e}}\big\langle
i\mathbf{\nabla} \cdot \left[\left(i\mathbf{\nabla}
\psi^{\dag}(x)\right)\sigma_{0}\psi(x)-\psi^{\dag}(x)\sigma_{0}
\left(i\mathbf{\nabla}\psi(x)\right)\right]\big\rangle \nonumber \\
&=& \big\langle i \partial_{x_0}\left(\psi^{\dag}(x)
\sigma_{3}\psi(x)\right)\big\rangle + \big\langle
\left(\xi_{\nabla}\psi^{\dag}(x)\right)\sigma_{0}\psi(x) -
\psi^{\dag}(x)\sigma_{0}\left(\xi_{\nabla}
\psi(x)\right)\big\rangle.
\end{eqnarray}
This then allows us to re-express the identity (\ref{eq:chargeSTI})
in terms of conserved current as follows
\begin{eqnarray}
\langle i\partial_{\mu}j^{c}_{\mu}(x)\rangle = \langle
\eta^{\dag}(x)\sigma_{3}\psi(x)\rangle - \langle
\psi^{\dag}(x)\sigma_{3}\eta(x)\rangle.
\end{eqnarray}

The current operators need to treated very carefully. In the
framework of local quantum field theory, the current operator
$j^{c}_{\mu}(x)$ is defined as the product of two spinor operators
$\psi^{\dag}(x)$ and $\psi(x)$ at one single point $x$. Calculations
based on such a definition often encounter short-distance
singularities \cite{PeskinbookSupp}. This is not surprising, since
the charge density is apparently divergent at one single space-time
point $x$. To avoid such singularities, it would be more suitable to
first define the charge density in a very small cube and then take
the volume of the cube to zero at the end of all calculations. The
singularities of current operators could be regularized by employing
the point-splitting technique. This technique was first proposed by
Dirac \cite{DiracSupp}, and later has been extensively applied to
deal with various field-theoretic problems \cite{PeskinbookSupp,
SchwingerSupp, Jackiw69Supp, Callan70Supp, Bardeen69Supp,
SchnablSupp, He01Supp}. In quantum gauge theories, such as QED and
QCD, the manipulation of point-splitting destroys local gauge
invariance, thus one needs to introduce a Wilson line to maintain
the gauge invariance of correlation functions. Moreover, the Lorentz
invariance may be explicitly broken. Fortunately, no such concerns
exist in EPI systems.

In order to derive WTIs, we only need to split the position vector
$\mathbf{x}$ into two close but separate points $\mathbf{x}$ and
$\mathbf{x}'$. The time $t$ need not be split. One might insist in
splitting $t$ into $t$ and $t'$, but the limit $t \rightarrow t'$
can be taken at any time. For notational simplicity, we shall use
the symbols $x$ and $x'$. Then we re-write the composite current
operators as follows
\begin{eqnarray}
\psi^{\dag}(x)\sigma_{0}\xi_{\nabla}\psi(x) -
\xi_{\nabla}\psi^{\dag}(x)\sigma_{0}\psi(x) &\rightarrow&
\psi^{\dag}(x) \sigma_{0}\xi_{\nabla_{x'}}\psi(x') -
\xi_{\nabla_{x}}\psi^{\dag}(x)\sigma_{0}\psi(x')
\nonumber \\
&=& (\xi_{\nabla_{x'}}-\xi_{\nabla_{x}})
\left(\psi^{\dag}(x)\sigma_{0}\psi(x')\right).\label{pst}
\end{eqnarray}
We will perform various field-theoretic calculations by making use
of this expression of current operator and take the limit $x
\rightarrow x'$ after all the calculations are completed.

The identity of Eq.~(\ref{eq:chargeSTI}) can be written as
\begin{eqnarray}
i \partial_{x_0} \langle (\psi^{\dag}(x)\sigma_{3}\psi(x))\rangle
-\lim_{x'\rightarrow x}(\xi_{\nabla_{x'}} - \xi_{\nabla_{x}})
\langle \psi^{\dag}(x)\sigma_{0}\psi(x')\rangle = \langle
\eta^{\dag}(x)\sigma_{3}\psi(x)\rangle - \langle
\psi^{\dag}(x)\sigma_{3}\eta(x)\rangle.\label{eq:chargeSTIsplitted}
\end{eqnarray}
As the next step, we perform functional derivatives
$\frac{\delta}{\delta\eta(z)}$ and
$\frac{\delta}{\delta\eta^{\dag}(y)}$ in order to both sides of this
equation. The calculational procedure has already been demonstrated
in Sec.~\ref{Sec:WTIES}. Analytic calculations show that
\begin{eqnarray}
&&i \partial_{x_0} \langle \psi^{\dag}(x)\sigma_{3}\psi(x)\psi(y)
\psi^{\dag}(z)\rangle -\lim_{x'\rightarrow x}(\xi_{\nabla_{x'}} -
\xi_{\nabla_{x}}) \langle \psi^{\dag}(x)\sigma_{0}\psi(x')\psi(y)
\psi^{\dag}(z)\rangle \nonumber \\
&=& \delta(y-x)\sigma_{3}G(x-z) - G(y-x)\sigma_{3}\delta(x-z).
\label{eq:chargewtispace}
\end{eqnarray}

For $\sigma_{i}=\sigma_{0}$. the transformation
Eq.~(\ref{eq:u1spin}) corresponds to another conserved current
$j^{s}_{\mu}(x) \equiv (j^{s}_{t}(x),\mathbf{j}^{s}(x))$, where
\begin{eqnarray}
j^{s}_{t}(x) &=& \psi^{\dag}(x)\sigma_{0}\psi(x), \\
\mathbf{j}^{s}(x) &=& \frac{1}{2m_{e}}\left[\left(i\mathbf{\nabla}
\psi^{\dag}(x)\right)\sigma_{3}\psi(x) - \psi^{\dag}(x)\sigma_{3}
\left(i\mathbf{\nabla}\psi(x)\right)\right]. \label{eq:scurrent}
\end{eqnarray}
Again, we see that $j^{s}_{t}(x)$ and $\mathbf{j}_{s}(x)$ are also
defined by two different $2\times 2$ matrices. One might notice
that, both $j^{c}_{t}(x)$ and $\mathbf{j}^{s}(x)$ involve
$\sigma_{3}$, and both $j^{s}_{t}(x)$ and $\mathbf{j}^{c}(x)$
involve $\sigma_{0}$. The current of Eq.~(\ref{eq:scurrent}) is also
conserved, namely $i\partial_{\mu}\langle j_{\mu}^{s}\rangle =0$, in
the absence of external sources, corresponding to spin conservation.
The generic identity Eq.~(\ref{eq:STInambu}) becomes
\begin{eqnarray}
\langle i \partial_{x_0}(\psi^{\dag}(x)\sigma_{0}\psi(x))\rangle +
\langle \xi_{\nabla}\psi^{\dag}(x)\sigma_{3}\psi(x) -
\psi^{\dag}(x)\sigma_{3}\xi_{\nabla}\psi(x)\rangle = \langle
\eta^{\dag}(x)\sigma_{0}\psi(x)\rangle - \langle
\psi^{\dag}(x)\sigma_{0}\eta(x)\rangle,\label{eq:spinSTI}
\end{eqnarray}
which can be re-written using the conserved current $j^{s}_{\mu}(x)$
as
\begin{eqnarray}
\langle i\partial_{\mu}j^{s}_{\mu}(x)\rangle = \langle
\eta^{\dag}(x)\sigma_{0}\psi(x)\rangle - \langle
\psi^{\dag}(x)\sigma_{0}\eta(x)\rangle.
\end{eqnarray}
The identity of Eq.~(\ref{eq:spinSTI}) can be written as
\begin{eqnarray}
i \partial_{x_0} \langle (\psi^{\dag}(x)\sigma_{0}\psi(x))\rangle
-\lim_{x'\rightarrow x}(\xi_{\nabla_{x'}} - \xi_{\nabla_{x}})
\langle \psi^{\dag}(x)\sigma_{3}\psi(x')\rangle = \langle
\eta^{\dag}(x)\sigma_{0}\psi(x)\rangle - \langle
\psi^{\dag}(x)\sigma_{0}\eta(x)\rangle.\label{eq:chargeSTIsplitted}
\end{eqnarray}
Perform functional derivatives $\frac{\delta}{\delta\eta(z)}$ and
$\frac{\delta}{\delta\eta^{\dag}(y)}$ in order to both sides of this
equation gives rise to
\begin{eqnarray}
&&i \partial_{x_0} \langle \psi^{\dag}(x)\sigma_{0}\psi(x)\psi(y)
\psi^{\dag}(z)\rangle -\lim_{x'\rightarrow x}(\xi_{\nabla_{x'}} -
\xi_{\nabla_{x}}) \langle \psi^{\dag}(x)\sigma_{3}\psi(x')\psi(y)
\psi^{\dag}(z)\rangle \nonumber \\
&=& \delta(y-x)\sigma_{0}G(x-z) - G(y-x)\sigma_{0}\delta(x-z).
\label{eq:spinwtispace}
\end{eqnarray}

To handle the identities given by Eq.~(\ref{eq:chargewtispace}) and
Eq.~(\ref{eq:spinwtispace}), we find it appropriate to define two
current vertex functions $\Gamma_{3}$ and $\Gamma_{0}$ as follows
\begin{eqnarray}
\langle \psi^{\dag}(x)\sigma_{i}\psi(x)\psi(y)\psi^{\dag}(z)\rangle
= -\int d\zeta d\zeta'
G(y-\zeta)\Gamma_{i}(\zeta-x,x-\zeta')G(\zeta'-z),
\label{eq:gammaidefinition}
\end{eqnarray}
where $\sigma_{0}$ corresponds to $\Gamma_{0}$ and $\sigma_{3}$
corresponds to $\Gamma_{3}$. If $x$ is splitted into two separate
points $x$ and $x'$, the above definition is changed into
\begin{eqnarray}
\langle \psi^{\dag}(x)\sigma_{i}\psi(x')\psi(y)\psi^{\dag}(z)\rangle
= \int d\zeta d\zeta'
G(y-\zeta)\Gamma_{i}(\zeta-x,x'-\zeta')G(\zeta'-z).
\label{eq:gammaidefinitionsplited}
\end{eqnarray}

The term $i \partial_{x_0}\langle \psi^{\dag}(x)\sigma_{i}\psi(x)
\psi(y)\psi^{\dag}(z)\rangle$ of Eq.~(\ref{eq:chargewtispace}) can
be treated
in the same way as what we have done in calculations
shown in Eq.~(\ref{eq:Fouriertransform1es}). It is straightforward
to show that
\begin{eqnarray}
i \partial_{x_0}\langle \psi^{\dag}(x)\sigma_{i}\psi(x) \psi(y)
\psi^{\dag}(z)\rangle &=& - \int d\zeta d\zeta' G(y-\zeta)
i\frac{\partial}{\partial
{x_{0}}}\Gamma_{i}(\zeta-x,x-\zeta')G(\zeta'-z)
\nonumber \\
&=& \int \frac{d^3p}{(2\pi)^3}\frac{d^3q}{(2\pi)^3}
G(p+q)q_{0}\Gamma_{i}(q,p)G(p) e^{-i(p+q)(y-x)}e^{-ip(x-z)}.
\label{eq:Fouriertransformt}
\end{eqnarray}
The term $\lim_{x'\rightarrow x}(\xi_{\nabla_{x'}} -
\xi_{\nabla_{x}})\langle
\psi^{\dag}(x)\sigma_{i}\psi(x')\psi(y)\psi^{\dag}(z)\rangle$ should
be treated more carefully. According to
Eq.~(\ref{eq:gammaidefinitionsplited}), we carry out Fourier
transformation as follows
\begin{eqnarray}
&&\lim_{x'\rightarrow x}(\xi_{\nabla_{x'}}-\xi_{\nabla_{x}})\langle
\psi^{\dag}(x)\sigma_{i}\psi(x')\psi(y)\psi^{\dag}(z)\rangle \nonumber \\
&=&-\lim_{x'\rightarrow x}(\xi_{\nabla_{x'}}-\xi_{\nabla_{x}})\int
d\zeta d\zeta' G(y-\zeta)\Gamma_{i}(\zeta-x,x'-\zeta')G(\zeta'-z)
\nonumber \\
&=&-\lim_{x'\rightarrow x}(\xi_{\nabla_{x'}}-\xi_{\nabla_{x}})\int
d\zeta d\zeta' \int\frac{d^3p_{1}}{(2\pi)^3} G(p_{1})
e^{-ip_{1}(y-\zeta)}\int\frac{d^3q}{(2\pi)^3} \frac{d^3p}{(2\pi)^3}
e^{-i(p+q)(\zeta-x)}\Gamma_{i}(q,p)e^{-ip(x'-\zeta')} \nonumber \\
&&\times \int\frac{d^3p_{2}}{(2\pi)^3}G(p_{2})e^{-ip_{2}(\zeta'-z)}
\nonumber \\
&=&-\lim_{x'\rightarrow x}(\xi_{\nabla_{x'}}-\xi_{\nabla_{x}}) \int
\frac{d^3p}{(2\pi)^3}\frac{d^3q}{(2\pi)^3}\left[\int
\frac{d^3p_{1}}{(2\pi)^3}\frac{d^3p_{2}}{(2\pi)^3}d\zeta d\zeta'
e^{i\left(p_{1}-(p+q)\right)\zeta}e^{i(p-p_{2})\zeta'}\right]
G(p_{1})\Gamma_{i}(q,p)G(p_{2}) \nonumber \\
&& \times e^{i(p+q)x-ipx'-ip_{1}y+ip_{2}z} \nonumber \\
&=&-\lim_{x'\rightarrow x}(\xi_{\nabla_{x'}}-\xi_{\nabla_{x}}) \int
\frac{d^3p}{(2\pi)^3}\frac{d^3q}{(2\pi)^3}\left[\int
\frac{d^3p_{1}}{(2\pi)^3}\frac{d^3p_{2}}{(2\pi)^3}(2\pi)^6
\delta\left(p_{1}-(p+q)\right)\delta(p-p_{2})\right]
G(p_{1})\Gamma_{i}(q,p)G(p_{2})\nonumber \\
&& \times e^{i(p+q)x-ipx'-ip_{1}y+ip_{2}z}
\nonumber \\
&=&-\lim_{x'\rightarrow x}(\xi_{\nabla_{x'}}-\xi_{\nabla_{x}})\int
\frac{d^3p}{(2\pi)^3}\frac{d^3q}{(2\pi)^3} G(p+q)\Gamma_{i}(q,p)
G(p)e^{i(p+q)x-ipx'-i(p+q)y+ipz}
\nonumber \\
&=& \int \frac{d^3p}{(2\pi)^3}\frac{d^3q}{(2\pi)^3}
G(p+q)\left(\xi_{\mathbf{p}}-\xi_{\mathbf{p+q}}\right)
\Gamma_{i}(q,p) G(p)e^{-i(p+q)(y-x)}e^{-ip(x-z)}.
\label{eq:Fouriertransformnabla}
\end{eqnarray}
The right-hand sides of Eq.~(\ref{eq:chargewtispace}) and
Eq.~(\ref{eq:spinwtispace}) can be treated as follows
\begin{eqnarray}
\delta(y-x)\sigma_{3}G(x-z)-G(y-x)\sigma_{3}\delta(x-z) &=& \int
\frac{d^3p}{(2\pi)^3}\frac{d^3q}{(2\pi)^3}
\left[\sigma_{3}G(p)-G(p+q)\sigma_{3}\right]
e^{-i(p+q)(y-x)-ip(x-z)}, \label{eq:Fouriertransformgpsigma3}\\
\delta(y-x)\sigma_{0}G(x-z)-G(y-x)\sigma_{3}\delta(x-z) &=& \int
\frac{d^3p}{(2\pi)^3} \frac{d^3q}{(2\pi)^3} \left[\sigma_{0}G(p) -
G(p+q)\sigma_{0}\right]e^{-i(p+q)(y-x)-ip(x-z)}.
\label{eq:Fouriertransformgpsigma0}
\end{eqnarray}
Based on the results of
Eqs.~(\ref{eq:Fouriertransformt}-\ref{eq:Fouriertransformgpsigma0}),
we now obtain two identities
\begin{eqnarray}
q_{0}G(p+q)\Gamma_{3}(q,p)G(p) - \left(\xi_{\mathbf{p+q}} -
\xi_{\mathbf{p}}\right)G(p+q)\Gamma_{0}(q,p)G(p) &=&
\sigma_{3}G(p)-G(p+q)\sigma_{3}, \\
q_{0}G(p+q)\Gamma_{0}(q,p)G(p) - \left(\xi_{\mathbf{p+q}} -
\xi_{\mathbf{p}}\right) G(p+q)\Gamma_{3}(q,p)G(p) &=&
\sigma_{0}G(p)-G(p+q)\sigma_{0}.
\end{eqnarray}
They can be readily re-written in more compact forms
\begin{eqnarray}
q_{0}\Gamma_{3}(q,p)-\left(\xi_{\mathbf{p+q}} -
\xi_{\mathbf{p}}\right) \Gamma_{0}(q,p) &=& G^{-1}(p+q)\sigma_{3} -
\sigma_{3}G^{-1}(p), \label{eq:chargewti} \\
q_{0}\Gamma_{0}(q,p) - \left(\xi_{\mathbf{p+q}} -
\xi_{\mathbf{p}}\right)\Gamma_{3}(q,p) &=& G^{-1}(p+q)\sigma_{0} -
\sigma_{0}G^{-1}(p).\label{eq:spinwti}
\end{eqnarray}
The WTIs given by Eq.~(\ref{eq:chargewti}) and
Eq.~(\ref{eq:spinwti}) come from the symmetries defined by
Eq.~(\ref{eq:u1charge}) and Eq.~(\ref{eq:u1spin}), respectively.
These two WTIs involve two unknown current vertex functions, namely
$\Gamma_{3}(q,p)$ and $\Gamma_{0}(q,p)$. After solving these two
self-consistently coupled identities, we can determine
$\Gamma_{3}(q,p)$ and $\Gamma_{0}(q,p)$ using $q_{0}$,
$\xi_{\mathbf{p+q}} - \xi_{\mathbf{p}}$, and $G(p)$.

The above two WTIs have already been derived in
Ref.~\cite{Liu2021Supp} in the case of pure EPI system. In the
present work, the model contains an additional Coulomb interaction
between electrons. Would the Coulomb interaction change such WTIs?
No changes at all. This is because the partition function describing
pure EPI and the one describing the interplay between EPI and
Coulomb interaction preserve the same U(1) symmetries defined by
Eq.~(\ref{eq:u1symmetrynambu}).

\section{Revisiting the WTI of Engelsberg and Schrieffer \label{sec:wtiessplitted}}

The point-splitting technique plays a crucial role in the derivation
of WTIs given by Eq.~(\ref{eq:chargewti}) and
Eq.~(\ref{eq:spinwti}). What would one obtain if this technique is
combined with the derivation of the WTI presented in
Ref.~\cite{Schrieffer1963Supp}. According to the analysis of
Sec.~\ref{Sec:WTIES}, we know that the following identity holds
\begin{eqnarray}
&&\frac{\delta}{\delta\eta^{\dag}(y)}\frac{\delta}{\delta\eta(z)}
\langle i \partial_{x_0} j_{0}(x) + i\mathbf{\nabla} \cdot
\mathbf{j}(x)\rangle \nonumber \\
&=& \langle i \partial_{x_0} j_{0}(x)\Psi(y)\Psi^{\dag}(z)\rangle +
\langle i \mathbf{\nabla}\cdot
\mathbf{j}(x)\Psi(y)\Psi^{\dag}(z)\rangle \nonumber \\
&=& i \partial_{x_0}\big\langle \Psi^{\dag}(x)\Psi(x)\Psi(y)
\Psi^{\dag}(z)\big\rangle + \big\langle\big[\left(\xi_{\nabla}
\Psi^{\dag}(x)\right)\Psi(x) - \Psi^{\dag}(x)\left(\xi_{\nabla}
\Psi(x)\right)\big]\Psi(y) \Psi^{\dag}(z)\big\rangle \nonumber \\
&=& \delta(y-x)G(x-z) - G(y-x)\delta(x-z).
\label{eq:partialjmupsipsidages}
\end{eqnarray}
Splitting $x$ into $x$ and $x'$ allows us to make the following
replacement
\begin{eqnarray}
\big\langle\big[\left(\xi_{\nabla} \Psi^{\dag}(x)\right)\Psi(x) -
\Psi^{\dag}(x) \left(\xi_{\nabla} \Psi(x)\right)\big]\Psi(y)
\Psi^{\dag}(z)\big\rangle \Rightarrow -\lim_{x\rightarrow
x'}(\xi_{\nabla_{x'}} - \xi_{\nabla_{x}})\langle
\Psi^{\dag}(x)\Psi(x')\Psi(y) \Psi^{\dag}(z)\rangle,
\end{eqnarray}
which then leads to
\begin{eqnarray}
&&i \partial_{x_0} \langle \Psi^{\dag}(x)\Psi(x)\Psi(y)
\Psi^{\dag}(z)\rangle -\lim_{x'\rightarrow x}(\xi_{\nabla_{x'}} -
\xi_{\nabla_{x}})\langle \Psi^{\dag}(x)\Psi(x')\Psi(y)
\Psi^{\dag}(z)\rangle \nonumber \\
&=& \delta(y-x)G(x-z) - G(y-x)\delta(x-z).
\label{eq:partialfourspinoressplitted}
\end{eqnarray}
It is now only necessary to define one single scalar function
${\tilde \Gamma}_{0}$ as
\begin{eqnarray}
\langle \Psi^{\dag}(x)\Psi(x)\Psi(y) \Psi^{\dag}(z)\rangle = -\int
d\xi d\xi' G(y-\xi){\tilde \Gamma}_{0}(\xi-x,x-\xi') G(\xi'-z),
\label{eq:tildegamma0defi}
\end{eqnarray}
which becomes
\begin{eqnarray}
\langle \Psi^{\dag}(x)\Psi(x')\Psi(y) \Psi^{\dag}(z)\rangle = -\int
d\xi d\xi' G(y-\xi){\tilde \Gamma}_{0}(\xi-x,x'-\xi') G(\xi'-z)
\label{eq:tildegamma0defisplitted}
\end{eqnarray}
after splitting $x$ into $x$ and $x'$. Then
Eq.~(\ref{eq:tildegamma0defi}) and
Eq.~(\ref{eq:tildegamma0defisplitted}) can be substituted into
Eq.~(\ref{eq:partialfourspinoressplitted}), yielding the following
identity
\begin{eqnarray}
&&\int d\xi d\xi' G(y-\xi)\big[i \partial_{x_0}{\tilde \Gamma}_{0}
(\xi-x, x-\xi') -\lim_{x'\rightarrow x}(\xi_{\nabla_{x'}} -
\xi_{\nabla_{x}}){\tilde \Gamma}_{0}(\xi-x,x'-\xi')\big] G(\xi'-z)
\nonumber \\
&=& G(y-x)\delta(x-z) -\delta(y-x)G(x-z).
\label{eq:wtirealspace}
\end{eqnarray}
Fourier transformation changes this identity into
\begin{eqnarray}
q_{0}{\tilde \Gamma}_{0}(q,p) - \left(\xi_{\mathbf{p+q}} -
\xi_{\mathbf{p}}\right){\tilde \Gamma}_{0}(q,p) =
G^{-1}(p+q)-G^{-1}(p).\label{eq:wtiessplitted}
\end{eqnarray}
Making use of the electron dispersion $\xi_{\mathbf{p}} =
\frac{\mathbf{p}^{2}}{2m_{e}}-\mu_{\mathrm{F}}$, one finds
\begin{eqnarray}
\left(\xi_{\mathbf{p+q}}-\xi_{\mathbf{p}}\right){\tilde \Gamma}_{0}
(q,p) = \mathbf{q}\cdot \frac{2\mathbf{p}+\mathbf{q}}{2m_{e}}
{\tilde \Gamma}_{0}(q,p).
\end{eqnarray}
Comparing to Eq.~(\ref{eq:chargewties}), we see that the vector
function ${\tilde {\mathbf \Gamma}}(q,p)$ studied in
Ref.~\cite{Schrieffer1963Supp} can be connected to ${\tilde
\Gamma}_{0}(q,p)$ via the relation
\begin{eqnarray}
{\tilde {\mathbf \Gamma}}(q,p) = \frac{2\mathbf{p} +
\mathbf{q}}{2m_{e}}{\tilde \Gamma}_{0}(q,p).
\end{eqnarray}
Using Eq.~(\ref{eq:wtiessplitted}), the scalar function ${\tilde
\Gamma}_{0}(q,p)$ can be determined by the full electron propagator
$G(p)$ via the relation
\begin{eqnarray}
{\tilde \Gamma}_{0}(q,p) = \frac{G^{-1}(p+q)-G^{-1}(p)}{q_{0} -
\left(\xi_{\mathbf{p+q}} - \xi_{\mathbf{p}}\right)}.
\end{eqnarray}

\end{widetext}


\begin{thebibliography}{99}

\bibitem{Schrieffer64}
J. R. Schrieffer, \emph{Theory of Superconductivity} (Taylor and
Francis, 1964).

\bibitem{Scalapino}
D. J. Scalapino, in \emph{Superconductivity}, edited by R. D. Parks
(Marcel Dekker, Inc. New York, 1969).

\bibitem{Allen}
P. B. Allen and B. Mitrovi\'{c}, \emph{Theory of Superconducting
$T_{c}$}, Solid State Physics, Vol.37 (Academic Press, Inc., 1982).

\bibitem{Carbotte}
J. P. Carbotte, Properties of boson-exchange superconductors, Rev.
Mod. Phys. {\bf 62}, 1027 (1990).

\bibitem{Marsiglio20}
F. Marsiglio, Eliashberg theory: A short review, Ann. Phys. {\bf
417}, 168102 (2020).

\bibitem{Migdal}
A. Migdal, Interaction between electrons and lattice vibrations in a
normal metal, Sov. Phys. JETP {\bf 7}, 996 (1958).

\bibitem{Eliashberg}
G. M. Eliashberg, Interactions between electrons and lattice
vibrations in a superconductor, Sov. Phys. JETP {\bf 11}, 696
(1960).

\bibitem{Tolmachev}
V. V. Tolmachev, Logarithmic criterion for superconductivity, Dokl.
Akad. Nauk SSSR {\bf 140}, 563 (1961).

\bibitem{Morel62}
P. Morel and P. W. Anderson, Calculation of the superconducting
state parameters with retarded electron-phonon interaction, Phys.
Rev. {\bf 125}, 1263 (1962).

\bibitem{Schooley64}
J. F. Schooley, W. R. Hosler, and M. L. Cohen, Superconductivity in
semiconducting SrTiO$_{3}$, Phys. Rev. Lett. {\bf12}, 474 (1964).

\bibitem{Fernandes20}
M. N. Gastiasoro, J. Ruhman, and R. M. Fernandes, Anisotropic
superconductivity mediated by ferroelectric fluctuations in cubic
systems with spin-orbit coupling, Ann. Phys. {\bf 417}, 168107
(2020).

\bibitem{Sham83}
H. Rietschel and L. J. Sham, Role of electron Coulomb interaction in
superconductivity, Phys. Rev. B {\bf 28}, 5100 (1983).

\bibitem{Richardson97}
C. F. Richardson and N. W. Ashcroft, Effective electron-electron
interactions and the theory of superconductivity, Phys. Rev. B {\bf
55}, 15130 (1997).

\bibitem{Prokefev22}
T. Wang, X. Cai, K. Chen, B. V. Svistunov, and N. V. Prokof'ev,
Origin of the Coulomb pseudopotential, Phys. Rev. B {\bf 107},
L140507 (2022).

\bibitem{Katsnelson22}
M. Simonato, M. I. Katsnelson, and M. R\"{o}sner, Revised
Tolmachev-Morel-Anderson pseudopotential for layered conventional
superconductors with nonlocal Coulomb interaction, Phys. Rev. B {\bf
108}, 064513 (2023).

\bibitem{Liu21}
G.-Z. Liu, Z.-K. Yang, X.-Y. Pan, and J.-R. Wang, Towards exact
solutions for the superconducting $T_{c}$ induced by electron-phonon
interaction, Phys. Rev. B {\bf103}, 094501 (2021).

\bibitem{Pan21}
X.-Y. Pan, Z.-K. Yang, X. Li, and G.-Z. Liu, Nonperturbative
Dyson-Schwinger equation approach to strongly interacting Dirac
fermion systems, Phys. Rev. B {\bf 104}, 085141 (2021).

\bibitem{Yang22}
Z.-K. Yang, X.-Y. Pan, and G.-Z. Liu, A non-perturbative study of
the interplay between electron-phonon interaction and Coulomb
interaction in undoped graphene, J. Phys.:Condens. Matter {\bf 35},
075601 (2022).

\bibitem{Nambu60}
Y. Nambu, Quasi-particles and gauge invariance in the theory of
superconductivity, Phys. Rev. {\bf 117}, 648 (1960).

\bibitem{Johnstoncp}
P. M. Dee, J. Coulter, K. G. Kleiner, and S. Johnston, Relative
importance of nonlinear electron-phonon coupling and vertex
corrections in the Holstein model, Communications Phys. {\bf 3}, 145
(2020).

\bibitem{Itzykson}
C. Itzykson and J.-B. Zuber, \emph{Quantum Field Theory},
(McGraw-Hill Inc. 1980).

\bibitem{Peskin}
M. E. Peskin and D. V. Schroeder, \emph{An Introduction to Quantum
Field Theory} (CRC Press, 2018).

\bibitem{Engelsberg}
S. Engelsberg and J. R. Schrieffer, Coupled electron-phonon system,
Phys. Rev. {\bf 131}, 993 (1963).

\bibitem{supplementary}
A detailed derivation of the two WTIs used in our calculations is
presented in the Supplementary Material. The Supplementary Material
also contains Refs.~[25-31].

\bibitem{Dirac}
P. A. M. Dirac, Discussion of the infinite distribution of electrons
in the theory of the positron, Proc. Camb. Phil. Soc. {\bf 30}, 150
(1934).

\bibitem{Schwinger}
J. Schwinger, On gauge invariance and vacuum polarization, Phys.
Rev. {\bf 82}, 664 (1951).

\bibitem{Jackiw69}
R. Jackiw and K. Johnson, Anomalies of the axial-vector current,
Phys. Rev. {\bf 182}, 1459 (1969).

\bibitem{Bardeen69}
W. A. Bardeen, Anomalous Ward identities in spinor field theories,
Phys. Rev. {\bf 184}, 1848 (1969).

\bibitem{Callan70}
C. G. Callan, S. Coleman, and R. Jackiw, A new improved
energy-momentum tensor, Ann. Phys. {\bf 59}, 42 (1970).

\bibitem{Schnabl}
J. Novotny and M. Schnabl, Point-splitting regularization of
composite operators and anomalies, Fortschr. Phys. {\bf 48}, 253
(2000).

\bibitem{He01}
H. He, F. C. Khanna, and Y. Takahashi, Transverse Ward-Takahashi
identity for the fermion-boson vertex in gauge theories, Phys. Lett.
B {\bf 480}, 222 (2000).

\bibitem{AGDbook}
A. A. Abrikosov, L. P. Gor'kov, and I. Y. Dzyaloshinskii,
\emph{Quantum Field Theoretical Methods in Statistical Physics}
(Pergamon Press Inc., 1965).

\bibitem{Chubukov04}
A. Abanov and A. V. Chubukov, Anomalous scaling at the quantum
critical point in itinerant antiferromagnets, Phys. Rev. Lett. {\bf
93}, 255702 (2004).

\bibitem{Metlitski10}
M. A. Metlitski and S. Sachdev, Quantum phase transitions of metals
in two spatial dimensions. I. Ising-nematic order, Phys. Rev. B {\bf
82}, 075127 (2010).

\bibitem{Liu19}
P.-L. Zhao and G.-Z. Liu, Absence of emergent supersymmetry in
superconducting quantum critical points in Dirac and Weyl
semimetals, npj Quantum Materials {\bf 4}, 37 (2019).

\bibitem{Torroba20}
J. Aguilera Damia, M. Solis, and G. Torroba, How non-Fermi liquids
cure their infrared divergences, Phys. Rev. B {\bf 102}, 045147
(2020).

\bibitem{Xue12}
Q.-Y. Wang, Z. Li, W.-H. Zhang, Z.-C. Zhang, J.-S. Zhang, W. Li, H.
Ding, Y.-B. Ou, P. Deng, K. Chang, J. Wen, C.-L. Song, K. He, J.-F.
Jia, S.-H. Ji, Y.-Y. Wang, L.-L. Wang, X. Chen, X.-C. Ma, and Q.-K.
Xue, Interface-induced high-temperature superconductivity in single
unit-Cell FeSe films on SrTiO$_{3}$, Chin. Phys. Lett. {\bf 29},
037402 (2012).

\bibitem{Shen14}
J. J. Lee, F. T. Schmitt, R. G. Moore, S. Johnston, Y.-T. Cui, W.
Li, M. Yi, Z. K. Liu, M. Hashimoto, Y. Zhang, D. H. Lu, T. P.
Devereaux, D.-H. Lee, and Z.-X. Shen, Interfacial mode coupling as
the origin of the enhancement of $T_{c}$ in FeSe films on
SrTiO$_{3}$, Nature (London) {\bf 515}, 245 (2014).

\bibitem{Lee15}
D.-H. Lee, What makes the $T_c$ of FeSe/SrTiO$_{3}$ so high, Chin.
Phys. {\bf 24}, 117405 (2015).

\bibitem{Gorkov16}
L. P. Gor'kov, Peculiarities of superconductivity in the
single-layer FeSe/SrTiO$_{3}$ interface, Phys. Rev. B {\bf 93},
060507(R) (2016).

\bibitem{Yao16}
Z.-X. Li, F. Wang, D.-H. Lee, and H. Yao, What makes the $T_c$ of
monolayer FeSe on SrTiO$_{3}$ so high: a sign-problem-free quantum
Monte Carlo study, Sci. Bull. {\bf 61}, 925 (2016).

\bibitem{Xiang12}
Y.-Y. Xiang, F. Wang, D. Wang, Q.-H. Wang, and D.-H. Lee,
High-temperature superconductivity at the FeSe/SrTiO$_{3}$
interface, Phys. Rev.B {\bf 86}, 134508 (2012).

\bibitem{Xing14}
B. Li, Z. W. Xing, G. Q. Huang, and D. Y. Xing, Electron-phonon
coupling enhanced by the FeSe/SrTiO$_{3}$ interface, J. Appl. Phys.
{\bf 115}, 193907 (2014).

\bibitem{Johnston-NJP2016}
L. Rademaker, Y. Wang, T. Berlijn, and S. Johnston, Enhanced
superconductivity due to forward scattering in FeSe thin films on
SrTiO$_{3}$ substrates, New J. Phys. {\bf 18}, 022001 (2016).

\bibitem{Dolgov17}
M. L. Kuli\'{c} and O. V. Dolgov, The electron-phonon interaction
with forward scattering peak is dominant in high $T_c$
superconductors of FeSe films on SrTiO$_{3}$ (TiO$_{2}$), New J.
Phys. {\bf 19}, 013020(2017).

\bibitem{Opp-PRB2018}
A. Aperis and P. M. Oppeneer, Multiband full-bandwidth anisotropic
Eliashberg theory of interfacial electron-phonon coupling and
high-$T_{c}$ superconductivity in FeSe/SrTiO$_{3}$, Phys. Rev. B
{\bf 97}, 060501(R) (2018).

\bibitem{Opp-PRB2021}
F. Schrodi, A. Aperis and P. M. Oppeneer, Induced odd-frequency
superconducting state in vertex-corrected Eliashberg theory, Phys.
Rev. B {\bf 104}, 174518 (2021).

\bibitem{Gongxg15}
Y. Xie, H.-Y. Cao, Y. Zhou, S. Chen, H. Xiang, and X.-G. Gong,
Oxygen vacancy induced flat phonon mode at FeSe/SrTiO$_{3}$
interface, Sci. Rep. {\bf 5}, 10011 (2015).

\bibitem{Nakanishi}
N. Nakanishi, Ward-Takahashi identities in quantum field theory with
spontaneously broken symmetry, Prog. of Theore. Phys. {\bf 51}, 1183
(1974).

\bibitem{Yanagisawa}
T. Yanagisawa, Nambu-Goldstone bosons characterized by the order
parameter in spontaneous symmetry breaking, J. Phys. Soc. Jap. {\bf
86}, 104711 (2017).

\bibitem{Anderson63}
P. W. Anderson, Plasmons, gauge invariance, and mass, Phys. Rev.
{\bf 130}, 439 (1963).

%\bibitem{Higgs64}
%P. W. Higgs, Phys. Rev. Lett. {\bf 13}, 508 (1964).

\bibitem{Yanase21}
T. Kitamura, T. Yamashita, J. Ishizuka, A. Daido, and Y. Yanase,
Superconductivity in monolayer FeSe enhanced by quantum geometry,
Phys. Rev. Research {\bf 4}, 023232 (2022).
\end{thebibliography}

\begin{thebibliography}{99}

\bibitem{Liu2021Supp}
G.-Z. Liu, Z.-K. Yang, X.-Y. Pan, and J.-R. Wang, Towards exact
solutions for the superconducting T$_{c}$ induced by electron-phonon
interaction, Phys. Rev. B {\bf 103}, 094501 (2021).

\bibitem{PeskinbookSupp}
M. E. Peskin and D. V. Schroeder, \emph{An Introduction to Quantum
Field Theory} (CRC Press, 2018).

\bibitem{ItzyksonSupp}
C. Itzykson and J.-B. Zuber, \emph{Quantum Field Theory},
(McGraw-Hill Inc. 1980).

\bibitem{Nambu60Supp}
Y. Nambu, Quasi-particles and gauge invariance in the theory of
superconductivity, Phys. Rev. {\bf 117}, 648 (1960).

\bibitem{Schrieffer1963Supp}
S. Engelsberg and J.R. Schrieffer, Coupled electron-phonon System,
Phys. Rev. 131, 993 (1963).

\bibitem{DiracSupp}
P. A. M. Dirac, The quantum theory of the electron, Proc. Camb.
Phil. Soc. {\bf 30}, 150 (1934).

\bibitem{SchwingerSupp}
J. Schwinger, On gauge invariance and vacuum polarization, Phys.
Rev. {\bf 82}, 664 (1951).

\bibitem{Jackiw69Supp}
R. Jackiw and K. Johnson, Anomalies of the axial-vector current,
Phys. Rev. {\bf 182}, 1459 (1969).

\bibitem{Callan70Supp}
C. G. Callan, S. Coleman, and R. Jackiw, A new improved
energy-momentum tensor, Ann. Phys. {\bf 59}, 42 (1970).

\bibitem{Bardeen69Supp}
W. A. Bardeen, Anomalous Ward Identities in Spinor Field Theories,
Phys. Rev. {\bf 184}, 1848 (1969).

\bibitem{SchnablSupp}
J. Novotny and M. Schnabl, Point-splitting regularization of
composite operators and anomalies, Fortschr. Phys. {\bf 48}, 253
(2000).

\bibitem{He01Supp}
H. He, F. C. Khanna, and Y. Takahashi, Transverse Ward-Takahashi
identity for the fermion-boson vertex in gauge theories, Phys. Lett.
B {\bf 480}, 222 (2000).

\bibitem{Johnston-NJP2016Supp}
L. Rademaker, Y. Wang, T. Berlijn, and S. Johnston, Enhanced
superconductivity due to forward scattering in FeSe thin films on
SrTiO$_{3}$ substrates, New J. Phys. {\bf 18}, 022001 (2016).

\end{thebibliography}
\end{document}